\documentclass{article}

\usepackage{PRIMEarxiv}

\usepackage[utf8]{inputenc}
\usepackage[T1]{fontenc}
\usepackage{hyperref}
\usepackage{url}
\usepackage{booktabs}
\usepackage{amsfonts}
\usepackage{amsmath}
\usepackage{amssymb}
\usepackage{nicefrac}
\usepackage{microtype}
\usepackage{fancyhdr}
\usepackage{graphicx}
\usepackage{enumitem}
\usepackage{multirow}
\usepackage{xcolor}
\usepackage{subcaption}
\usepackage{pifont}
\usepackage{authblk}
\newcommand{\cmark}{\ding{51}}
\newcommand{\xmark}{\ding{55}}
\graphicspath{{figures/}}

\newcommand{\datasetname}{FoundationalASSIST}
\newcommand{\datasetnamenospace}{FoundationalASSIST} 

\title{FoundationalASSIST: An Educational Dataset for \\ Foundational Knowledge Tracing and \\ Pedagogical Grounding of LLMs}

\author[1]{Eamon Worden$^{*}$}
\author[2]{Cristina Heffernan}
\author[1]{Neil Heffernan}
\author[3]{Shashank Sonkar}

\affil[1]{Worcester Polytechnic Institute, Worcester, MA}
\affil[2]{The ASSISTments Foundation, Auburn, MA}
\affil[3]{University of Central Florida, Orlando, FL}

\begin{document}
\maketitle
\begingroup
\renewcommand\thefootnote{}
\footnotetext{$^{*}$Corresponding author: Eamon Worden. Emails: \texttt{elworden@wpi.edu}, \texttt{cristina.heffernan@assistments.org}, 
\texttt{nth@wpi.edu}, 
\texttt{shashank.sonkar@ucf.edu}.}
\endgroup

\begin{abstract}
Can Large Language Models understand how students learn? As LLMs are deployed for adaptive testing and personalized tutoring, this question becomes urgent---yet we cannot answer it with existing resources. Current educational datasets provide only question identifiers and binary correctness labels, rendering them opaque to LLMs that reason in natural language. We address this gap with \textbf{FoundationalASSIST}, the first English educational dataset providing the complete information needed for research on LLMs in education: full question text, actual student responses (not just right/wrong), records of which wrong answers students chose, and alignment to Common Core K-12 standards. These 1.7 million interactions from 5,000 students enable research directions that were previously impossible to pursue, from fine-tuning student models to analyzing misconception patterns.
To demonstrate the dataset's utility, we evaluate four frontier models (GPT-OSS-120B, Llama-3.3-70B, Qwen3-Next-80B variants) on two complementary task families: \textbf{Knowledge Tracing}, testing whether LLMs can predict student performance on questions, and the exact answer a student will give; and \textbf{Pedagogical Grounding}, testing whether LLMs understand the properties that make assessment items effective.
Our evaluation reveals significant gaps in current LLM capabilities. Every model barely achieves a trivial baseline on knowledge tracing: simply predicting ``correct'' performs nearly as well as any LLM. All models fall below random chance on item discrimination, indicating that LLMs do not understand what makes one problem more diagnostic than another. Models do show competence at judging relative difficulty (up to 68.6\%), but this partial success only highlights the gaps elsewhere. These results establish that substantial advances are needed before LLMs can reliably support personalized learning at scale. We release FoundationalASSIST to support progress on these foundational challenges.

\end{abstract}


\section{Introduction}
\label{sec:introduction}

The promise of personalized education has motivated research in educational technology for decades \cite{anderson1995cognitive,graesser2012autotutor,team2024learnlm}. Achieving this requires two complementary capabilities. First, \textit{knowledge tracing}: modeling what students know and predicting what they will struggle with next \cite{corbett1995knowledge,piech2015deep}, enabling adaptive systems to select optimal practice problems and identify at-risk students before they fall behind. Second, \textit{pedagogical grounding}: understanding the properties that make educational content effective, from item difficulty and discrimination \cite{baker2001basics} to which wrong answers reveal common misconceptions \cite{gurung2023common,chen2020impact,llmcogscience}. Together, these capabilities form the foundation for intelligent tutoring systems \cite{anderson1985intelligent} that can adapt to individual learners.

Large Language Models (LLMs) have offered new opportunities for both knowledge tracing and pedagogical grounding \cite{sonkar2023deduction,feng2025reasoning,scarlatos2025smart,pedalign,scarlatos2025exploring}. Modern LLMs demonstrate remarkable capabilities in reasoning, explanation, and natural language understanding, which present new opportunities for solving educational problems and benefiting students within intelligent tutoring systems. Commercial deployments already show LLMs generating practice problems, explaining solution strategies, and engaging students in Socratic dialogue \cite{carnegie2024livehint}. This raises a fundamental question: \textit{Can LLMs actually understand how students learn, or do they merely excel at tasks that happen to be linguistically similar \cite{sonkar2025turing}?} 

The answer matters because LLMs are increasingly being deployed for tasks that require an understanding of student cognition. Computerized adaptive testing with LLMs \cite{cheng2024towards} depends on predicting which students will succeed on which problems---and on selecting items that efficiently distinguish between ability levels. LLM-based tutors that provide personalized feedback \cite{reddig2025generating} need cognitive models of student thinking to target misconceptions effectively. Automated distractor generation requires anticipating which wrong answers students will find tempting, yet LLMs are not adept at anticipating common errors \cite{liu2025llms,feng2024distractor}. Item difficulty estimation \cite{scarlatos2025smart} requires LLMs to be capable of reasoning how students at different ability levels will approach a problem. Each of these applications requires foundational capabilities that, despite their importance, have never been rigorously evaluated.

We cannot currently answer this question, and the reason is simple: existing educational datasets are incompatible with the main capabilities of language models--natural language. The most widely used knowledge tracing benchmarks, including the ASSISTments datasets \cite{ASSISTmentsData20092010ASSISTment,ASSISTmentsData201213School,2017ASSISTmentsDatamining}, EdNet \cite{choi2020ednet}, and Junyi Academy \cite{pojen2020junyi}, represent each student interaction as a tuple of identifiers with a binary correctness label. A question ID conveys no semantic content; an LLM cannot reason about \textit{why} a student might struggle with problem 47829 with knowledge concept 21 without knowing what that problem asks. Binary correctness labels are equally incompatible with LLMs. When a student selects distractor ``B'' on a multiple-choice problem, that choice often reveals a specific misconception. When they write ``1/4'' instead of ``0.25'', it suggests conceptual understanding but unfamiliarity with decimal notation. These diagnostic signals vanish when correctness is reduced to zeros and ones.

This paper addresses the dataset gap with \textbf{\datasetnamenospace{}}, the first English-language educational dataset that provides the textual richness LLMs require. Built using data from the ASSISTments platform \cite{heffernan2014assistments}, \datasetnamenospace{} preserves what made ASSISTments the gold standard for math education research while adding content LLMs need to reason about student learning. The dataset includes complete question text for 3,400 mathematics problems, enabling models to reason about content and difficulty. It records actual student responses for 1.7 million interactions: which specific option(s) students selected on 0.5 million multiple-choice problems, and what numerical or textual answers they wrote for the 1.2 million fill-in-the-blank responses. It provides the full set of answer choices for each multiple-choice item, enabling analysis of distractor effectiveness and common misconceptions. Furthermore, all problems are from the Illustrative Mathematics curriculum \cite{illustrative_mathematics_curriculum}, which aligns every problem with the Common Core State Standards \cite{CCSSMath2010}, the official U.S. K-12 mathematics curriculum, complete with codes that encode grade level and knowledge concept(s).

These properties enable research questions that were simply impossible to investigate with prior datasets. Cognitive student modeling, where we predict the exact response a student will give rather than just whether they will be correct, requires actual responses. Distractor analysis, identifying which wrong answers trap students most often, requires knowing which distractors students selected. The closest prior work, XES3G5M \cite{liu2023xes3g5m}, provides question text but only binary correctness labels, and is available only in Chinese. To our knowledge, \datasetnamenospace{} is the first educational dataset in any language to provide actual student responses and distractor selection data alongside complete question text.

To demonstrate what the dataset enables and to establish baselines for future research, we evaluate LLMs on tasks spanning both foundational capabilities. The \textbf{Knowledge Tracing} tasks test the prediction capabilities of LLMs on two key facets of students' performance: whether a student will answer a specific question correctly, and what exact response they will produce. The \textbf{Pedagogical Grounding} tasks test whether LLMs understand the psychometric properties that make assessment items effective: which of two problems is more difficult, which better discriminates between high- and low-ability students, and which distractors students choose most and least often. Together, these consist of six tasks that probe the foundational capabilities required for effective implementation of LLMs into intelligent tutoring systems.

We evaluated four state-of-the-art LLMs on this benchmark: GPT-OSS-120B \cite{agarwal2025gpt}, Llama-3.3-70B-Instruct \cite{touvron2023llama}, and two variants of Qwen3-Next-80B (Instruct and Thinking) \cite{qwen2025qwen3}. The results suggest LLMs still require significant development, prompt engineering, or fine-tuning before they perform well on these tasks. On knowledge tracing, every model barely achieves a trivial baseline. Simply predicting that students will answer correctly will result in a 51.3\% accuracy rate of our evaluation set. We found improvements over the trivial baseline were limited: the strongest model surpassed it by only five percentage points, while others achieved gains of merely one percentage point. The models also exhibit systematic optimistic bias: Llama-3.3-70B correctly predicts correctness with 85.4\% accuracy when students answer correctly, but only 12.6\% accuracy when they answer incorrectly. A tutor with this bias would systematically fail to recognize struggling students, a crucial component of all knowledge tracing algorithms.

The pedagogical grounding results reveal a further pedagogical gap. While models show reasonable competence at comparing item difficulty, achieving up to 80.0\% accuracy when problems differ substantially, they completely fail to understand item discrimination. Every model performs \textit{below random chance} on discrimination comparison, suggesting they do not grasp what makes one problem more diagnostic than another. This is not a minor limitation. Understanding discrimination is central to educational measurement; an item that fails to distinguish high-ability from low-ability students provides no useful information, regardless of its difficulty. Distractor prediction reveals a similar asymmetry: models achieve 47.9\% accuracy at identifying which wrong answer students choose most often, exceeding the 35.8\% random baseline, but fall 15 percentage points \textit{below} chance when predicting which errors students rarely select. LLMs have learned to recognize some common misconceptions but cannot distinguish likely from unlikely distractors.

Extended reasoning has mixed effects. The Qwen3-Next-80B ``Thinking'' variant, which generates explicit reasoning chains, achieves the highest discrimination accuracy at 46.9\%, though still below 50\%. Yet the same model achieves the lowest distractor prediction accuracy at 20.5\%. We believe explicit reasoning may help with abstract pedagogical judgments while introducing overthinking on more concrete predictions.

Just as the original ASSISTments datasets have been used for over a decade of knowledge tracing research, we intend \datasetnamenospace{} to serve as the foundational benchmark for evaluating LLMs in education. By establishing baselines and identifying specific tasks LLMs struggle with, we aim to focus research on the advances needed for LLMs to reliably support personalized learning. The remainder of this paper presents related work (Section~\ref{sec:related-work}), dataset details (Section~\ref{sec:dataset}), task definitions (Section~\ref{sec:tasks}), experimental setup (Section~\ref{sec:setup}), results and analysis (Section~\ref{sec:results}), discussion of implications and limitations (Section~\ref{sec:discussion}), and conclusions (Section~\ref{sec:conclusion}).

\section{Related Work}
\label{sec:related-work}

\datasetnamenospace{} contributes to three important areas of educational research: knowledge tracing, educational datasets, and the emerging application of LLMs to education. Each area has made substantial progress, yet each has left gaps that our work helps address.

\subsection{The Evolution of Knowledge Tracing}

Knowledge tracing attempts to model what skills students know, how well they know them, and how their knowledge of these skills changes as they practice over time. Corbett and Anderson's \textbf{Bayesian Knowledge Tracing} \cite{corbett1995knowledge} operationalized this using Hidden Markov Models, estimating a binary skill mastery through four parameters: initial knowledge, learning rate, guess probability, and slip probability. BKT remains influential decades later, but its assumptions, such as assuming that a student's mastery of a skill is binary, limit how it captures the nuanced ways students actually learn.

Deep learning offered a new approach to knowledge tracing. Piech et al.'s \textbf{Deep Knowledge Tracing} \cite{piech2015deep} demonstrated that recurrent neural networks could capture complex temporal patterns in student response sequences, outperforming BKT at the cost of added model complexity. This led to further architectural advancements: attention mechanisms \cite{ghosh2020context}, memory networks \cite{zhang2017dynamic}, and eventually transformer architectures \cite{pandey2019self}, each pushing prediction accuracy higher. Yet most knowledge tracing research has operated at the \textit{skill level}, predicting aggregate mastery of abstract knowledge components rather than performance on specific questions. To address this limitation, researchers began incorporating question-level representations—such as in Question-centric Deep Knowledge Tracing \cite{sonkar2020qdkt}—enabling models to predict performance on individual problems rather than abstract skills and problem identifiers alone.

Researchers have also moved beyond binary correctness toward richer response modeling. \textbf{Option Tracing} \cite{ghosh2021option} models which specific answer option a student selects on multiple-choice problems, which allows for more thorough analysis regarding specific misconceptions. \textbf{Open-ended Knowledge Tracing} \cite{liu2022open} extends the paradigm to free-response problems in computer science, using language models to encode student code submissions. Our benchmark continues this trajectory by asking whether LLMs can perform both binary problem correctness and cognitive (exact textual value of answer) prediction, leveraging their natural language understanding of question content in ways that ID-based models cannot.

\subsection{The Dataset Bottleneck}

Progress in knowledge tracing has been enabled and also constrained by the datasets available. The ASSISTments platform \cite{heffernan2014assistments} has produced several valuable datasets, including ASSIST 2009, 2012, 2015, and 2017. These datasets record millions of student interactions with mathematics problems and have become the standard benchmarks against which many knowledge tracing methods are evaluated.

But these datasets have a significant limitation that matters specifically for language model research: they contain only problem identifiers and binary correctness labels. The actual question text, the responses students wrote, and the specific distractors they selected are unavailable, typically due to licensing restrictions and data format conventions that made sense for ID-based models. An LLM confronting these datasets sees sequences of meaningless identifiers and binary outcomes; it cannot apply its language understanding capabilities because there is no language to understand. Other datasets, such as \textbf{EdNet} \cite{choi2020ednet}, offer impressive scale, with over 130 million interactions from a Korean educational platform. Yet it shares the same structural limitation.

The closest prior work to ours is \textbf{XES3G5M} \cite{liu2023xes3g5m}, which provides question text and hierarchical knowledge component annotations from a Chinese K-12 platform. This dataset enabled important research on language-aware knowledge tracing. However, XES3G5M still only records binary correctness, not what students actually answered or which distractors they selected. Moreover, containing Chinese-language data limits its applicability to English-based LLM research.

\datasetnamenospace{} addresses these limitations directly. We provide actual student responses for 1.7 million interactions, including the 1.2 million fill-in responses where students wrote their own answers. We record distractor selection data showing which specific wrong answer each student chose. We provide English question text that enables LLM natural language understanding. Further, as we curated the dataset to include only problems aligned to Common Core State Standards, providing a standardized K-12 curriculum framework with clear progression.

Table~\ref{tab:dataset-comparison} in Section~\ref{sec:dataset} summarizes how \datasetnamenospace{} compares with prior datasets.

Research has also examined how problem format affects learning. Gurung et al.~\cite{gurung2024multiple} studied trade-offs between multiple-choice and fill-in-the-blank problems in ASSISTments, finding that while multiple-choice problems scale more easily, fill-in problems may promote deeper learning. Because \datasetnamenospace{} includes both formats, it enables analysis of how LLMs perform on various tasks, such as knowledge tracing, across problem types.

\subsection{LLMs Enter Education}

Large Language Models are impacting educational technology. In intelligent tutoring, LLMs generate explanations \cite{worden2025scaling}, provide hints \cite{pardos2024chatgpt}, and engage students in Socratic dialogue \cite{fakour2025socratic,sonkar2023class}. Commercial systems like Khan Academy's Khanmigo \cite{Khanmigo2025} and Duolingo's AI \cite{DuolingoMax2025} features demonstrate that these capabilities can work at scale.

For question creation, LLMs can generate practice problems, produce varied question phrasings, and create distractors for multiple-choice items \cite{moore2024automated}. Creating high-quality practice materials is expensive, requiring experienced teachers, and LLMs present a significantly cheaper option, if they can generate effective, appropriately difficult, pedagogically sound questions and distractors.

Yet student modeling, a core challenge of knowledge tracing, and one which is integral to many features of intelligent tutors, remains underexplored. Solving problems correctly and predicting how students will solve them require different capabilities. The former asks ``what is the right answer?'' while the latter asks ``what will this student answer, given what I know about their learning history?'' This second question demands understanding of student cognition, common misconceptions, and how knowledge develops over time. Our benchmark directly tests whether LLMs possess this understanding. Research has shown LLMs doing knowledge tracing may struggle less with cold-start issues \cite{jung2024clst} and can be integrated into dialogue-based systems\cite{scarlatos2025exploring}, but the lack of natural-language centric datasets has limited research in this area.

Much of the existing LLM research has focused on improving mathematical problem-solving accuracy, relying on benchmark datasets such as GSM8K \cite{cobbe2021gsm8k} and FrontierMath \cite{glazer2024frontiermath} to determine the mathematical abilities of LLMs. Much of this work, however, evaluates LLMs primarily on their own performance—whether they can solve problems correctly or articulate high-quality explanations. Our benchmark instead treats student learning as the object of prediction: can an LLM anticipate what a particular student will answer, including systematic errors, given their learning history? This distinction separates reasoning about content from reasoning about learners.

\subsection{Psychometric Foundations}

Our pedagogical grounding tasks build on Item Response Theory \cite{lord2012applications,embretson2013item}, the psychometric framework that underlies modern educational assessment. IRT models the probability of a correct response as a function of student ability and item characteristics.

The two-parameter logistic model characterizes each item by its difficulty ($b$), the ability level required for a student to have a 50\% probability of success, and its discrimination ($a$), how effectively the item separates high-ability from low-ability students. A high-discrimination item has a steep probability curve: students slightly below the difficulty threshold almost always fail, while those slightly above almost always succeed. A low-discrimination item shows similar performance across ability levels, providing little diagnostic information regardless of its difficulty. Understanding difficulty is relatively intuitive, as harder problems require more knowledge or skill. Discrimination is subtler. A highly discriminating item is one where capable students consistently succeed, and less capable students consistently fail, typically because the item tests core concepts without introducing confounding factors like unclear wording or irrelevant knowledge requirements \cite{rush2016impact}.

Recent work has begun integrating LLMs with psychometric frameworks, such as IRT models. Scarlatos et al.~\cite{scarlatos2025smart} introduced SMART, using direct preference optimization to train LLMs as simulated students with instructed ability levels, then fitting synthetic responses to IRT models to predict item difficulty without real student data. Feng et al. \cite{feng2025reasoning} augment difficulty prediction with LLM-generated reasoning chains that decompose the cognitive steps each answer option requires.

These approaches typically require fine-tuning or specialized training, which would benefit heavily from large natural-language datasets. Further, it remains an open question whether pretrained LLMs are capable of understanding and predicting psychometric concepts from their training corpora. \datasetnamenospace{} provides the first English benchmark to test this systematically. Our pedagogical grounding tasks evaluate whether LLMs can compare item difficulty, identify discriminating items, and predict common misconceptions without additional training. As we will show, the results are mixed: LLMs have reasonable intuitions about difficulty, but perform notably worse at predicting whether items are discriminating.

\section{The \datasetnamenospace{} Dataset}
\label{sec:dataset}

\subsection{From ASSISTments to \datasetnamenospace{}}

The ASSISTments platform has contributed to knowledge tracing research by providing high-quality datasets for over a decade \cite{feng2009addressing,heffernan2014assistments}. ASSISTments provides K-12 students with mathematics practice, delivering immediate feedback and on-demand support as they work through problems. The ASSISTments datasets have been used as benchmarks in a variety of tasks, ranging from knowledge tracing to psychometric analysis. Prior releases shared a limitation that had little impact on traditional knowledge tracing but poses a major challenge for LLM-based methods: Question content and knowledge concepts were represented solely as IDs, and student responses were collapsed into binary correctness labels rather than retaining the original multiple-choice selections or textual answers.

\datasetnamenospace{} breaks from this tradition. Drawing on student interactions from January 2019 through July 2024, we preserve what prior releases omitted: complete problem text, the full set of answer options, and the actual responses students provided. A model evaluating this dataset does not see that student 4521 answered problem 892 incorrectly; it sees that the student was asked to find how many batches of spice mix can be made with 9 cups of chili powder when one batch requires $\frac{3}{4}$ cup, and that the student wrote ``27'' rather than the correct answer of ``12''. This difference transforms what LLMs can learn and what researchers can evaluate.

\subsection{Scale and Scope}

Table~\ref{tab:dataset-overview} summarizes the dataset.

\begin{table}[h]
\centering
\caption{Overview of the \datasetnamenospace{} dataset.}
\label{tab:dataset-overview}
\begin{tabular}{lr}
\toprule
\textbf{Statistic} & \textbf{Value} \\
\midrule
Total Student Interactions & 1,722,169 \\
Unique Students & 5,000 \\
Unique Problems & 3,395 \\
Knowledge Components (Skills) & 462 \\
Skill Annotations & 355,307 \\
Date Range & Jan 2019 -- Jul 2024 \\
Avg. Interactions per Student & 344.4 \\
Overall Correctness Rate & 61.5\% \\
\bottomrule
\end{tabular}
\end{table}

\datasetnamenospace{} includes 1.7 million learning interactions from 5,000 students working through 3,395 unique mathematics problems, with each student completing between 211 and 421 interactions (344 on average). The dataset was constructed to include a large number of problems per student, enabling a variety of tasks such as knowledge tracing, modeling misconceptions, and more.

\subsection{What the Data Contains}

The dataset comprises three files. The interaction data records each student-problem encounter: an anonymized student identifier, problem identifier, the student's actual response (an option letter for multiple-choice or a value for fill-in problems), binary correctness, the number of hints requested, whether the student revealed the answer before responding, and the timestamp of completion.

The problem content file describes each problem: its full text in natural language (with HTML and MathML for mathematical notation), problem type (multiple choice, fill-in-the-blank, and others), the correct answer, and the set of answer choices for multiple-choice items.

The skill mappings file connects problems to the curriculum. Each problem links to one or more Common Core State Standards, the official U.S. K-12 mathematics curriculum. A standard code like ``6.NS.C.6'' encodes both grade level (Grade 6) and content domain (Number System, Cluster C, Standard 6), along with a textual description ``Identify Opposites of Integers'', enabling analysis that aligns with how mathematics is actually taught.

\subsection{Problem Types and Their Challenges}

Table~\ref{tab:problem-types} shows the distribution of problem types.

\begin{table}[h]
\centering
\caption{Distribution of problem types by unique problems and student interactions.}
\label{tab:problem-types}
\begin{tabular}{lrrrr}
\toprule
\textbf{Problem Type} & \textbf{Problems} & \textbf{\%} & \textbf{Interactions} & \textbf{\%} \\
\midrule
Fill-in-the-blank(s) & 2,188 & 64.4\% & 1,199,593 & 69.7\% \\
Multiple Choice (select 1) & 795 & 23.4\% & 328,329 & 19.1\% \\
Multiple Choice (select all) & 380 & 11.2\% & 210,927 & 12.2\% \\
Order / Sort & 32 & 0.9\% & 14,761 & 0.9\% \\
\bottomrule
\end{tabular}
\end{table}

Fill-in-the-blank problems make up the majority of the dataset, comprising 64.4\% of unique problems and 69.7\% of interactions. For LLM evaluation, fill-in problems pose a novel challenge: predicting that a student will write ``27'' when the correct answer is ``12'' is far harder than predicting they will select option B. The remaining problems include standard multiple-choice (select one option), multi-select (choose all correct answers), and ordering questions (arrange items in sequence). While similar, each type has unique challenges that researchers will need to address when running their own analysis on the dataset.

\subsection{Difficulty and Baseline Performance}

The overall correctness rate of the dataset is 61.5\%. However, due to the computation cost of knowledge tracing with LLMs, we did not predict a student's performance on every question. Rather, we created a sample which included a correctness rate of 51.3\% to have a more balanced test set. A model that ignores student history and simply predicts ``correct'' for every problem achieves this accuracy without any student modeling. 

Accuracy varies predictably with problem type. Single-select multiple choice problems see 70.5\% accuracy, boosted by the opportunity for correct guessing. Fill-in problems have a correctness rate of 61.7\%, requiring students to generate rather than recognize answers. Multi-select and ordering problems hover around 47-48\%, where the complexity of response options suppresses accuracy even among capable students.

\subsection{What Makes \datasetnamenospace{} Different}

Four properties distinguish this dataset from prior knowledge tracing benchmarks and enable the evaluations we propose.

First, complete question text. Where prior datasets provide only identifiers, we include full problem statements. Consider this example: \textit{``A recipe for 1 batch of spice mix calls for $\frac{3}{4}$ cups of chili powder and $\frac{1}{4}$ cup of cumin. How many batches can be made with 9 cups of chili powder?''} An LLM can reason about what makes this problem challenging, which skills it requires, and what misconceptions might lead students astray. An LLM given only ``problem\_47829'' can do nothing of the sort.

Second, actual student responses. Rather than reducing the student work to binary correctness, we preserve what students wrote. For the problem above, there is diagnostic value in distinguishing a student who wrote ``12'' (correct) from one who wrote ``27'' (multiplied 9 by 3, ignoring the fraction) from one who wrote ``$\frac{1}{12}$'' (inverted the relationship). These errors reveal distinct misconceptions, and any model hoping to understand student cognition must be capable of understanding these misconceptions.

Third, distractor selection data. For multiple-choice problems, we record which specific wrong answer each student selected. This goes beyond binary correctness to reveal systematic error patterns. When 40\% of incorrect responses on a problem select option B while only 10\% select option D, the data suggest that option B embodies a common misconception worth analyzing. This enables our distractor prediction tasks, which test whether LLMs can anticipate which errors students will make. Further, by including the entire text associated with each option, LLMs can reason better about why each distractor is or is not selected by students.

Fourth, Common Core curriculum alignment, complete with knowledge concepts and the associated natural language descriptions. Every problem maps to one or more Common Core State Standards, providing a structured view of the K-12 mathematics curriculum. Researchers can analyze performance by grade level, track how models handle different mathematical domains in middle school mathematics, and align findings with how mathematics is actually taught in schools using ASSISTments.

\subsection{Preprocessing}

Problem text arrives with HTML markup and MathML notation for mathematical expressions. We provide the necessary code that converts this markup to more readable text, handling fraction notation, exponents, and special characters while preserving mathematical structure. The cleaned text is what LLMs receive in our benchmark evaluation.

\subsection{Comparison with Prior Datasets}

Table~\ref{tab:dataset-comparison} situates \datasetnamenospace{} among existing knowledge tracing benchmarks.

\begin{table}[h]
\centering
\caption{Comparison of \datasetnamenospace{} with existing educational datasets. \datasetnamenospace{} is the only dataset combining question text, actual student responses, and distractor selection data in English, enabling comprehensive LLM-based cognitive modeling.}
\label{tab:dataset-comparison}
\begin{tabular}{lrrrcccc}
\toprule
\textbf{Dataset} & \textbf{Interact.} & \textbf{Students} & \textbf{Avg/Stu} & \textbf{Q. Text} & \textbf{Resp.} & \textbf{Dist.} & \textbf{Eng.} \\
\midrule
ASSIST 2009 \cite{ASSISTmentsData20092010ASSISTment} & 347K & 4K & 82 & \xmark & \xmark & \xmark & \cmark \\
ASSIST 2012 \cite{ASSISTmentsData201213School} & 2.5M & 27K & 94 & \xmark & \xmark & \xmark & \cmark \\
ASSIST 2017 \cite{2017ASSISTmentsDatamining} & 943K & 686 & 1374 & \xmark & \xmark & \xmark & \cmark \\
EdNet \cite{choi2020ednet} & 131M & 784K & 167 & \xmark & \xmark & \xmark & \xmark \\
Junyi Academy \cite{pojen2020junyi} & 16M & 72K & 223 & \xmark & \xmark & \xmark & \xmark \\
Eedi \cite{wang2020diagnostic} & 20M & 125K & 160 & Partial & Partial& Partial & \cmark \\
XES3G5M \cite{liu2023xes3g5m} & 5.5M & 18K & 307 & \cmark & \xmark & \xmark & \xmark \\
\midrule
\textbf{\datasetnamenospace{} (Ours)} & \textbf{1.7M} & \textbf{5K} & \textbf{344} & \cmark & \cmark & \cmark & \cmark \\
\bottomrule
\end{tabular}
\end{table}

Some prior datasets are larger in raw scale. EdNet contains over 130 million interactions from 784,000 students; Junyi Academy and Eedi each provide 16 and 20 million interactions, respectively. But none provides the complete set of properties that LLM-based evaluation requires. The closest is Eedi--which provides an image rather than text of each question and distractor. However, the Eedi dataset is also limited to only multiple-choice problems. The table reveals a clear pattern: existing datasets do not satisfy requirements for LLM-based user modelling---question text, actual student responses, distractor selection data, and English language. XES3G5M offers question text but records only binary correctness and is available only in Chinese. Eedi provides partial question text in English but lacks actual responses. The ASSISTments datasets, while influential, provide neither question text nor response content.

\datasetnamenospace{} is the only dataset with all four properties. It includes 344 interactions per student on average---among the longest learning histories among released benchmarks---which provides the temporal information that knowledge tracing demands. This combination unlocks evaluations that were simply not possible before.

\subsection{Dataset Characteristics}

\begin{figure}[t]
    \centering
    \begin{subfigure}[b]{\textwidth}
        \includegraphics[width=\textwidth]{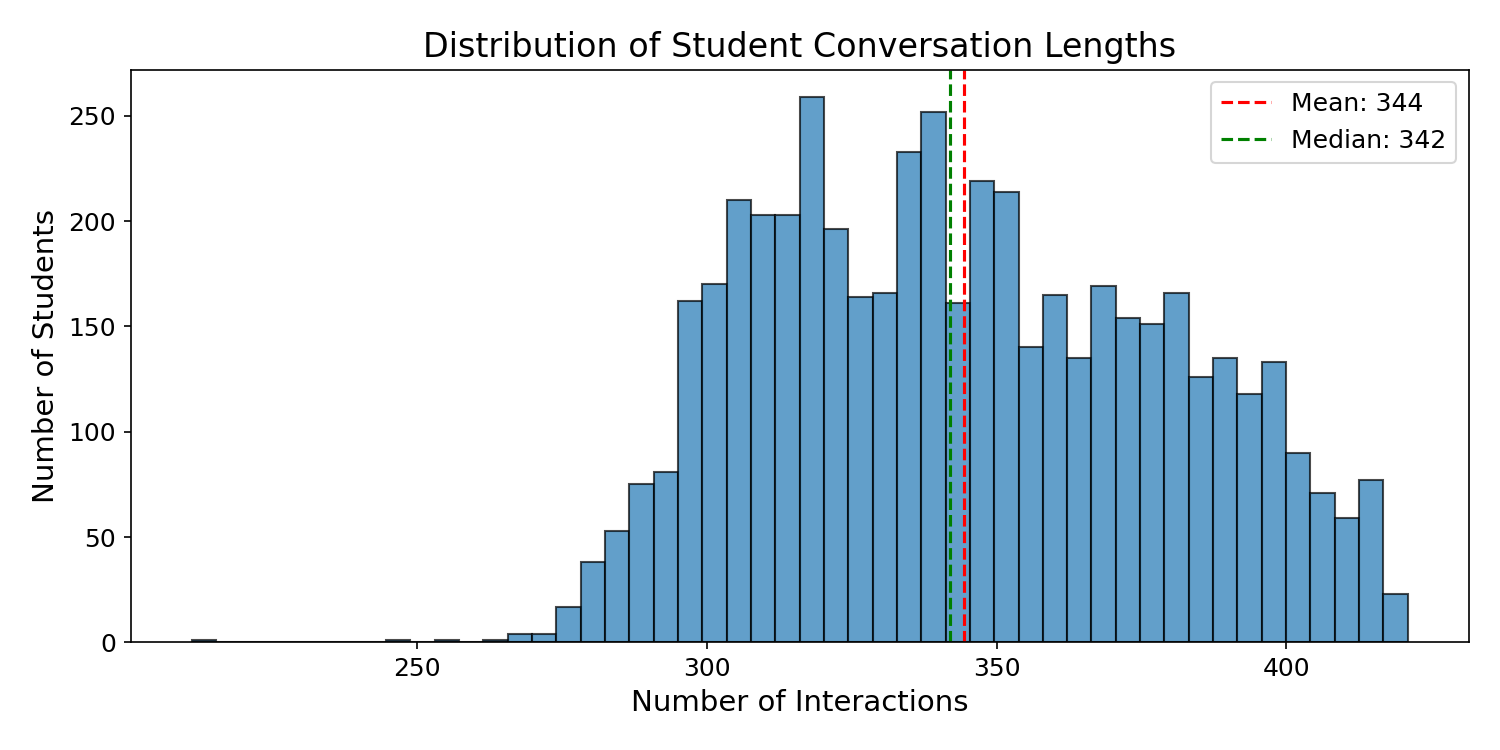}
        \caption{Distribution of interactions per student.}
        \label{fig:conv-lengths}
    \end{subfigure}

    \vspace{0.5em}

    \begin{subfigure}[b]{\textwidth}
        \centering
        \includegraphics[width=0.8\textwidth]{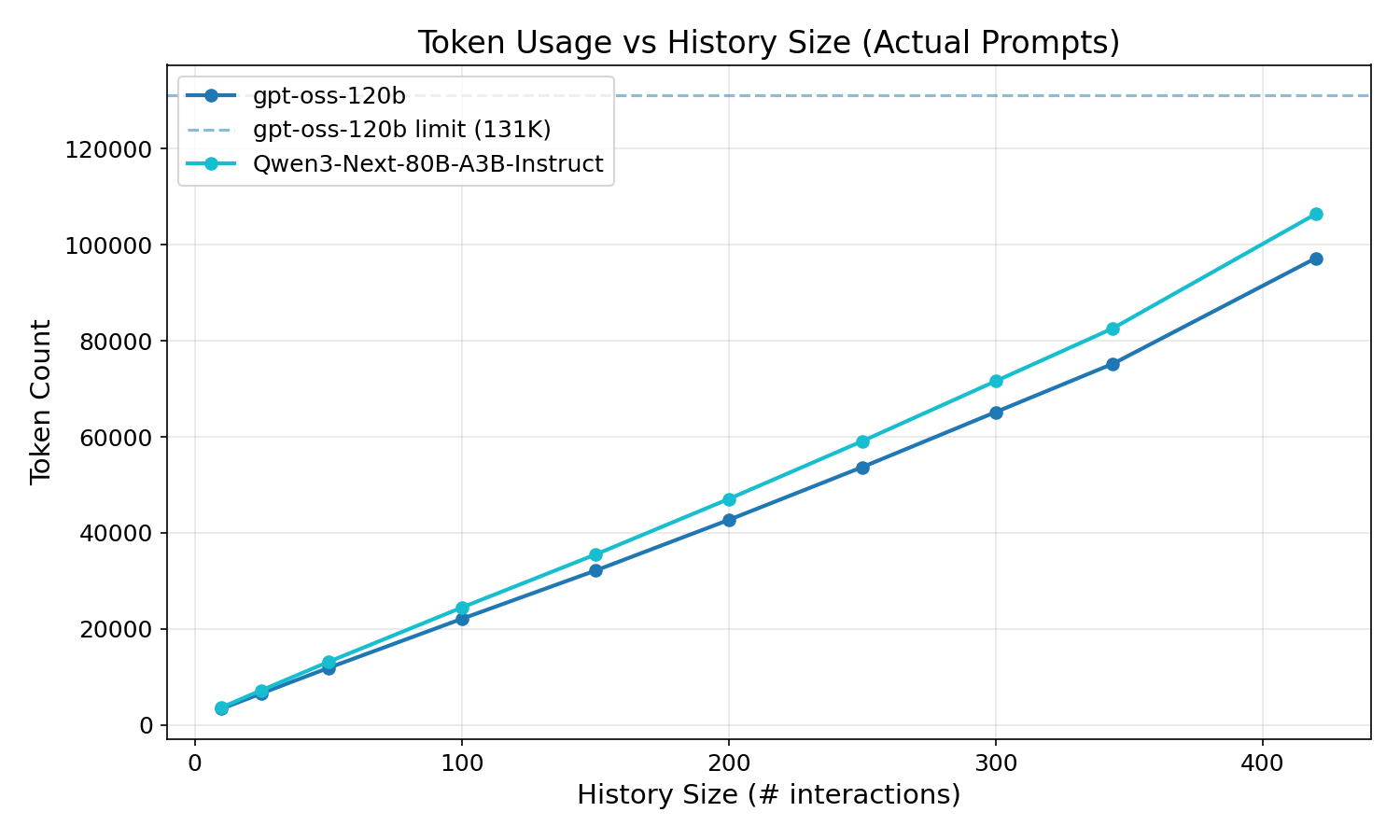}
        \caption{Token counts for student histories.}
        \label{fig:tokens}
    \end{subfigure}
    \caption{Student learning histories. (a) Students complete 211--421 interactions each, averaging 344. (b) Full histories approach 32K tokens, challenging LLM context limits.}
    \label{fig:student-activity}
\end{figure}

Figure~\ref{fig:student-activity} shows student learning trajectories. Each student completes 211--421 problems, with an average of 344 interactions. When encoded as prompts, these histories require tens of thousands of tokens, testing the limits of current LLM context windows.

\begin{figure}[t]
    \centering
    \begin{subfigure}[b]{\textwidth}
        \includegraphics[width=\textwidth]{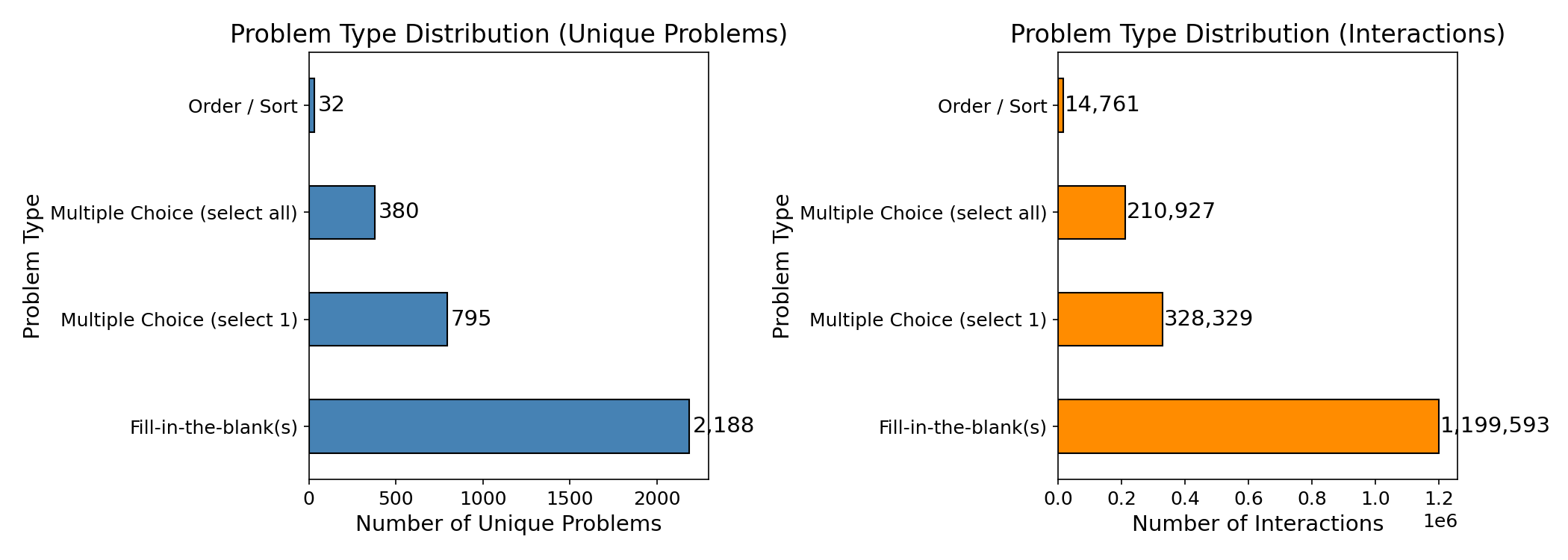}
        \caption{Problem types by interaction count.}
        \label{fig:problem-types-fig}
    \end{subfigure}

    \vspace{0.5em}

    \begin{subfigure}[b]{\textwidth}
        \includegraphics[width=\textwidth]{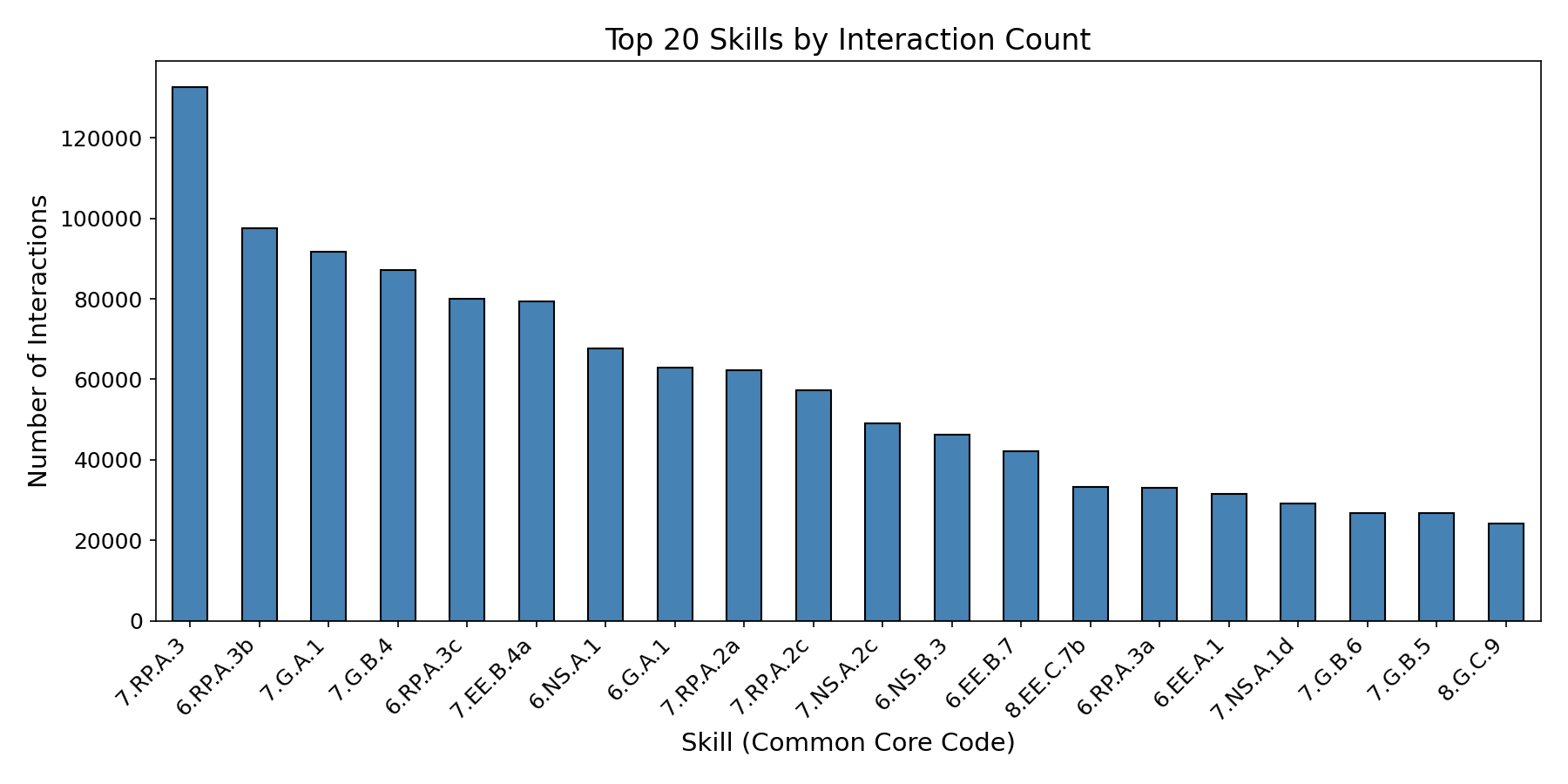}
        \caption{Common Core skill coverage.}
        \label{fig:skills}
    \end{subfigure}
    \caption{Content distribution. (a) Fill-in-the-blank problems dominate (70\% of interactions), posing a harder prediction challenge than multiple-choice. (b) Skills span all Common Core math domains from grades 3--8.}
    \label{fig:content-distribution}
\end{figure}

Figure~\ref{fig:content-distribution} shows the distribution of problem types and curriculum coverage. Fill-in-the-blank problems make up 69.7\% of interactions, requiring models to predict exact numerical or textual answers rather than selecting from options. The problems span all Common Core mathematics domains from grades 3--8; however is curated to be overwhelmingly comprised of problems for grades 6-8.

\section{Benchmark Tasks}
\label{sec:tasks}

In this section, we define the benchmark tasks that \datasetname enables. We propose two complementary evaluation tracks: \textbf{Knowledge Tracing} (Section~\ref{sec:kt-tasks}) and \textbf{Pedagogical Grounding} (Section~\ref{sec:pg-tasks}). We propose two knowledge tracing tasks and four pedagogical grounding tasks on which to evaluate the capabilities of LLMs in educational contexts.

\subsection{Knowledge Tracing Tasks}
\label{sec:kt-tasks}

The Knowledge Tracing track evaluates whether LLMs can model student learning and predict future performance. We define two tasks that share a common input format but differ in what the model must predict: binary correctness or exact student responses.

\subsubsection{Task Formulation}

For each student, we construct a learning history as a sequence of interactions:
\[
\mathcal{H} = \{(q_i, p_i, r_i, a_i, o_i, k_i, c_i, h_i, d_i, t_i)\}_{i=1}^{T}
\]
where for each interaction $i$: $q_i$ is the question text, $p_i$ is the problem type, $r_i$ is the student's actual response (option letter for multiple-choice, value for fill-in), $a_i$ is the correct answer(s), $o_i$ is the answer options (omitted for fill-in problems), $k_i$ is the set of knowledge components (Common Core skills), $c_i \in \{0, 1\}$ indicates correctness, $h_i$ is the number of hints used, $d_i$ indicates whether the student revealed the answer before responding, and $t_i$ is the timestamp.

Given history $\mathcal{H}_{1:T}$ and a new question $q_{T+1}$ with its skill labels $k_{T+1}$, problem type $p_{T+1}$, correct answer $a_{T+1}$, and answer options (if applicable) $o_{T+1}$, the model must predict how the student will perform. The student's future response $r_{T+1}$, hint usage $h_{T+1}$, and whether they revealed the answer $d_{T+1}$ are all withheld; these represent future actions the model must anticipate, which also impact whether a student was awarded full points ($c_{T+1}$).

\subsubsection{Task 1: Question-Level Knowledge Tracing}

The first task is the classical knowledge tracing formulation: given the student's history and the next question, predict whether they will answer correctly.

The model receives history $\mathcal{H}_{1:T}$ and new question $(q_{T+1}, k_{T+1}, p_{T+1}, a_{T+1}, o_{T+1})$, and outputs a binary prediction $\hat{c}_{T+1} \in \{0, 1\}$. This task matters because accurate prediction enables adaptive systems to select appropriately challenging problems and identify when students need additional support.

We evaluate using accuracy (proportion of correct binary predictions) and AUC-ROC (to account for class imbalance between correct and incorrect responses). The baseline of 51.3\% represents always predicting ``correct''; any model with genuine predictive power should exceed this.

\subsubsection{Task 2: Cognitive Student Modeling}

The second task is challenging but achievable for teachers \cite{gurung2023identification}, and the most challenging for LLMs: predict the student's exact response. For multiple-choice problems, this means predicting which option the student will select. For fill-in-the-blank problems, this means predicting the numerical or textual answer the student will write.

This task tests whether LLMs can model student cognition beyond binary correctness, as has been done in a number of prior works \cite{ghosh2021option, liu2022open}. When a student selects a particular wrong answer, that choice often reveals a specific misconception. When students write incorrect numerical answers, the errors frequently follow predictable patterns such as off-by-one errors, inverted operations, unit conversion mistakes. A model that understands student thinking should anticipate these patterns.

We evaluate using exact match accuracy for multiple-choice and numerical tolerance (relative error $<$ 1\%) for fill-in problems, reporting metrics separately by problem type. This task is substantially harder than binary prediction: random guessing yields approximately 25\% for 4-option multiple-choice but near-zero for fill-in problems with infinite answer spaces.

\subsubsection{Prompt Design}

We use structured prompts that present student history chronologically, followed by the target question. The prompt instructs the model to output predictions in JSON format for reliable parsing. A shortened version of the prompt is presented:

\begin{quote}
\small
\texttt{You are an educational AI assistant analyzing student learning. Below is a student's problem-solving history, showing each question they attempted and their response.}

\texttt{[Student History: chronological list of $(q_i, p_i, r_i, a_i, o_i, k_i, c_i, h_i, d_i, t_i)$]}

\texttt{Now predict how this student will perform on the following new question:}

\texttt{[New Question with answer options if MC]}

\texttt{Provide your predictions in JSON format:}\\
\texttt{\{"skill\_level": true/false, "question\_level": 0/1, "student\_answer": "..."\}}
\end{quote}

For knowledge tracing, we predicted skill-level mastery, question-level correctness, and exact answer. Due to the ability
of LLMs to understand surface-level constructs, such as problem correctness, but not latent-level constructs, such as
skill-mastery, we only included question-level correctness and exact answer predictions in our task evaluations. The prompt includes timestamps and hint/answer-reveal indicators, as these provide an important signal about response reliability. A correct answer after revealing the solution does not indicate true understanding, and while the student's answer may be correct, they would receive a score of 0 since they saw the answer.

\subsection{Pedagogical Grounding Tasks}
\label{sec:pg-tasks}

The Pedagogical Grounding track evaluates whether LLMs understand fundamental properties of educational assessments. Rather than modeling individual students, these tasks test whether LLMs have internalized the pedagogical knowledge that expert educators and psychometricians use when designing effective assessments.

\subsubsection{Item Response Theory Background}

Our ground truth for difficulty and discrimination comes from Item Response Theory \cite{lord2012applications, embretson2013item}, the standard psychometric framework for analyzing educational assessments. The two-parameter logistic (2PL) model characterizes each problem by two parameters:

\textbf{Difficulty} ($b$) represents the ability level at which a student has 50\% probability of answering correctly. Higher values indicate harder problems. Difficulty is relatively intuitive: a problem asking students to multiply three-digit numbers is harder than one asking them to add single digits.

\textbf{Discrimination} ($a$) captures how effectively the problem distinguishes between high-ability and low-ability students. High-discrimination problems sharply separate students by ability--students with an ability above a problem's difficulty almost always succeed, and those with an ability level below almost always fail. Low-discrimination problems show similar performance across ability levels, often because guessing, ambiguous wording, or irrelevant complexity obscures the underlying skill being measured \cite{rush2016impact}.

We compute IRT parameters using Bayesian inference (py-irt) on 1.7 million student responses across 2,548 problems, filtering for items with at least 50 responses to ensure reliable parameter estimates. This resulted in 2788 items included in our IRT model.

\subsubsection{Task 3: Difficulty Comparison}

Given two problems, the model must identify which is more difficult.

The model receives a problem pair $(q_A, q_B)$ with full question text and answer options, and outputs a binary choice: ``A'' if $q_A$ is harder, ``B'' otherwise. Ground truth is the problem with a higher IRT difficulty parameter.

This task matters because understanding relative difficulty is essential for adaptive learning. A tutor that cannot recognize difficulty differences may overwhelm struggling students with challenges beyond their reach or bore advanced learners with trivial exercises.

We sample pairs using stratification based on difficulty difference $|b_A - b_B|$: small (0.1-0.5), medium (0.5-1.0), and large ($>$1.0). We selected equal numbers from each stratum ensure evaluation across the full range of difficulty gaps. Random baseline is 50\%.

\subsubsection{Task 4: Discrimination Comparison}

Given two problems, the model must identify which better distinguishes high-ability from low-ability students.

The model receives a problem pair and outputs a binary choice based on which problem is more discriminating. Ground truth is the problem with a higher 2-pl model IRT discrimination parameter.

This task is inherently harder than difficulty comparison. Difficulty can often be estimated from surface features such as the complexity of numbers, the number of solution steps, required skills, and more. Discrimination depends on how the problem structure interacts with student cognition. Does the problem have distractors that trap low-ability students while high-ability students see through them? Does it test core understanding without confounding factors? These questions require reasoning about how students with different abilities would approach the problem, not just about the problem itself.

We sample pairs using stratification by discrimination difference, with strata of small (0.1-0.5) and medium (0.5-1.0). Random baseline is 50\%.

\subsubsection{Knowledge of Student Misconceptions}

Research in science education has established that teachers' ability to predict students' most common wrong answers is a significant predictor of student learning gains \cite{chen2020impact}. Chen et al.\ found that when teachers could identify which distractor students would most frequently select, their students showed greater improvement on those items. This capability, termed \textit{Knowledge of Student Misconceptions} (KOSM), is distinct from subject matter knowledge: knowing the right answer is not the same as understanding which wrong answers students find tempting and why.

This has important implications for AI usage in educational contexts. If LLMs are to serve as effective tutors or assessment tools, they should demonstrate a similar understanding of common student errors. A tutor that cannot anticipate misconceptions cannot proactively address them. Tasks 5 and 6 test whether LLMs possess this capability.

The baseline for predicting the exact answer a student will give on a multiple-choice problem was calculated using the expected probability of correctly guessing by chance, given the number of answer options. For a problem with $k$ options, the probability of a correct guess is $\frac{1}{k}$; we computed the overall baseline as the weighted mean of these probabilities across all multiple-choice problems. Due to a large number of problems with only two options, this results in a baseline of 41.3\%. For select-all problems, we assumed each permutation of selected options was equally likely, meaning for a problem with $k$ options, the probability of a correct guess is $\frac{1}{2^k-1}$. This gives a baseline of 2.9\%.

\subsubsection{Task 5: Most Common Distractor Prediction}

For a multiple-choice problem, the model must predict which wrong answer students choose most often.

The model receives a problem with question text, correct answer, and a list of distractors, then outputs the distractor letter predicted to be most frequently selected. Ground truth is the distractor with the highest observed selection frequency across all student responses.

Effective distractors are ``wrong for a reason'': they capture specific misconceptions that students commonly hold. For the problem ``What is $\frac{1}{2} + \frac{1}{3}$?'' with options including $\frac{5}{6}$ (correct), $\frac{2}{5}$ (adding numerators and denominators), and $\frac{1}{5}$ (less interpretable), predicting that students commonly choose $\frac{2}{5}$ demonstrates understanding of the widespread misconception that fractions are added by summing numerators and denominators separately.

Random baseline is $\frac{1}{n-1}$ where $n$ is the number of options, yielding approximately 35.8\% weighted across problems.

\subsubsection{Task 6: Least Common Distractor Prediction}

The model must predict which wrong answer students choose least often.

This mirrors Task 5 but asks which distractor is ``obviously wrong'' to most students. The least effective distractor typically does not correspond to any coherent misconception; it may be an implausible number or a clearly inappropriate answer format. Identifying these reveals whether LLMs understand which errors are plausible versus implausible. Ground truth and baseline are also analogous to Task 5.

\subsubsection{Prompt Design}

For comparison tasks, we present both problems and request a direct choice:

\begin{quote}
\small
\texttt{Compare these two math problems and determine which is [harder / more discriminating].}

\texttt{Problem A: [question text and options]}\\
\texttt{Problem B: [question text and options]}

\texttt{Answer with just the letter: A or B}
\end{quote}

For distractor tasks, we present the problem with its options:

\begin{quote}
\small
\texttt{For this multiple-choice problem, predict which wrong answer students are [most / least] likely to choose.}

\texttt{Question: [question text]}\\
\texttt{Options: A) ... B) ... C) ... D) ...}\\
\texttt{Correct Answer: [letter]}

\texttt{Which distractor is [most / least] common? Answer with just the letter.}
\end{quote}

These zero-shot prompts test inherent pedagogical understanding without task-specific training or examples.

\section{Experimental Setup}
\label{sec:setup}

\subsection{Models Evaluated}

We evaluate four state-of-the-art LLMs representing different organizations, architectures, and reasoning strategies.

\textbf{GPT-OSS-120B} \cite{openai2025gptoss} is OpenAI's 120-billion parameter open-weight model. It uses a mixture-of-experts architecture with 5.1 billion active parameters. This model has a maximum context window of 131,072 tokens.\newline
\textbf{Llama-3.3-70B-Instruct} \cite{touvron2023llama} is Meta's 70-billion parameter instruction-tuned model, designed for multilingual capability and general-purpose instruction following. It has a maximum context window of 128,000 tokens.\newline
\textbf{Qwen3-Next-80B-Instruct} \cite{qwen2025qwen3} is Alibaba's 80-billion parameter model using ultra-sparse mixture-of-experts with only 3 billion active parameters. All models in the Qwen3 family have a maximum context window of 262,144 tokens.\newline
\textbf{Qwen3-Next-80B-Thinking} \cite{qwen2025qwen3} is a reasoning-enhanced variant of the Qwen3 family that generates explicit reasoning chains before producing answers, which enables analysis of whether extended reasoning helps educational tasks.

All models are deployed locally using vLLM with tensor parallelism across multiple GPUs.

\subsection{Knowledge Tracing Experimental Design}

Our knowledge tracing evaluation simulates a realistic tutoring scenario: given a student's complete learning history, predict how they will perform on upcoming problems.

We sample 500 students from the dataset. For each student, we group interactions into bins of 10 consecutive problems. Beginning with the 50th problem a student attempts, we selected the first correct and incorrect problem in each bin, if available (if the bin was homogenous we used only the first problem in the bin). We then constructed the students problem solving history before each problem and prompted each model to predict performance on the sampled problems within the bin. This yields approximately 27,000 predictions per model, repeating until we ran out of problems for each student.

To avoid cold-start artifacts where minimal history provides insufficient signal, we require 50 interactions before beginning evaluation. These initial interactions serve as warmup: included in context to establish the student's profile, but not evaluated as predictions. While prior works suggest fine-tuning LLMs may be able to mitigate the cold start problems associated with knowledge tracing~\cite{jung2024clst}, we do not focus on that as one of our benchmarks.

Student histories can require a large number of tokens. With an average of 344 interactions per student and hundreds of tokens per problem, contexts approach 32,000 tokens. We use the maximum context window available for each model.

Inference uses vLLM for parallelism across four A100 GPUs, with prefix caching enabled. Sampling parameters follow each model's recommendations from their HuggingFace model cards: temperature 0.6-0.7, top-$p$ 0.8-0.95, and top-$k$ 20 where applicable. We set the maximum generation length to 32,768 tokens to accommodate extended reasoning. The batch size is 512 predictions per inference call.

\subsection{Pedagogical Grounding Experimental Design}

Pedagogical grounding evaluations require different setups. For difficulty and discrimination comparison, we analyze 2,548 problems with IRT parameters computed from the full response data. For distractor prediction, we use 236 multiple-choice problems, each of which had at least 10 responses and one wrong response.

Comparison tasks use 10,000 problem pairs sampled with stratification: equal numbers from small-difference, medium-difference, and large-difference strata, with a minimum difference threshold of 0.1 to ensure meaningful comparisons. All 236 multiple-choice problems with valid distractor data were included at least once in the analysis.

Inference configuration matches knowledge tracing, but shorter contexts (single problems with distractors or pairs of problems rather than full student histories), which enabled faster completion.

\subsection{Compute Resources}

All experiments use NVIDIA A100 GPUs (80GB). Knowledge tracing requires four GPUs per model for tensor parallelism, with each model completing inference in approximately 2 days. Pedagogical grounding takes 2-4 hours per model due to shorter contexts. Total compute is approximately 400 GPU-hours across all experiments.

\subsection{Evaluation Protocol}

All evaluations use zero-shot prompting: models receive task descriptions and inputs, but no examples or fine-tuning. This tests what LLMs understand from pretraining rather than what they can learn from task-specific supervision.

For reproducibility, we fix random seeds and use consistent sampling parameters per model. Model responses are parsed using regex patterns to extract structured predictions from free-form text. Predictions that cannot be parsed are marked as incorrect, ensuring that evaluation penalizes both incorrect answers and failure to follow formatting instructions.

\section{Results and Analysis}
\label{sec:results}


We present results from evaluating four state-of-the-art LLMs on the \datasetnamenospace{} benchmark. All experiments use zero-shot prompting to assess inherent capabilities rather than task-specific learning.

Table~\ref{tab:models-summary} summarizes the models evaluated and their key characteristics.

\begin{table}[h]
\centering
\caption{Summary of evaluated models.}
\label{tab:models-summary}
\begin{tabular}{lcccc}
\toprule
\textbf{Model} & \textbf{Parameters} & \textbf{Active Parameters} & \textbf{Organization} & \textbf{Type} \\
\midrule
GPT-OSS-120B \cite{agarwal2025gpt} & 120B & 5.1B & OpenAI & Instruct \\
Llama-3.3-70B-Instruct \cite{touvron2023llama} & 70B & 70B & Meta & Instruct \\
Qwen3-Next-80B-Instruct \cite{yang2025qwen3} & 80B & 3B & Alibaba & Instruct \\
Qwen3-Next-80B-Thinking \cite{yang2025qwen3} & 80B & 3B & Alibaba & Reasoning \\
\bottomrule
\end{tabular}
\end{table}

The ``Thinking'' variant of Qwen3 generates explicit reasoning chains before providing answers, enabling investigation of whether extended deliberation improves pedagogical judgment.

We first present Knowledge Tracing results (Section~\ref{sec:kt-results}), analyzing performance on question-level prediction and cognitive modeling. We then present Pedagogical Grounding results (Section~\ref{sec:pg-results}), examining difficulty comparison, discrimination comparison, and distractor prediction tasks.

\subsection{Knowledge Tracing Results}
\label{sec:kt-results}

\subsubsection{Barely Above Baseline}

We found that each LLM was barely able to achieve the knowledge tracing baseline of 51.3\%, suggesting that without fine-tuning or prompt engineering, they are not well-suited for this task.

\begin{table}[h]
\centering
\caption{Knowledge Tracing results. FKT = Foundational Knowledge Tracing (question-level binary prediction). Cognitive = exact answer prediction. Evaluation set baseline = 51.3\% (always predict correct).}
\label{tab:kt-main-results}
\begin{tabular}{lcccc}
\toprule
\textbf{Model} & \textbf{N} & \textbf{FKT Acc.} & \textbf{AUC-ROC} & \textbf{Cognitive Acc.} \\
\midrule
GPT-OSS-120B & 27,715 & 56.2\% & 0.559 & 41.2\% \\
Llama-3.3-70B-Instruct & 27,715 & 52.3\% & 0.512 & 44.3\% \\
Qwen3-Next-80B-Instruct & 27,618 & 54.0\% & 0.539 & 35.5\% \\
Qwen3-Next-80B-Thinking & 27,715 & 55.8\% & 0.555 & 38.5\% \\
\midrule
\textit{Baseline} & --- & 51.3\% & 0.500 & --- \\
\bottomrule
\end{tabular}
\end{table}

A model that ignores student history entirely and predicts ``correct'' for every question achieves 51.3\% accuracy on the evaluation portion of our dataset. The best-performing models, GPT-OSS-120B and Qwen3-Next-80B-Thinking, reach 56.2\% and 55.8\%, respectively, beating the baseline by just around 5 percentage points. This is not a competitive result; it suggests that current LLMs, despite their impressive reasoning capabilities, struggle to leverage student interaction histories for performance prediction. Llama-3.3 achieved just 52.3\% accuracy, which is marginally above the baseline accuracy.

\subsubsection{Extended Reasoning Helps, But Not Enough}

The ``Thinking'' variant of Qwen3 generates explicit reasoning chains before making predictions. Comparing directly to its Instruct counterpart, extended reasoning improves FKT accuracy from 54.0\% to 55.8\% and AUC-ROC from 0.539 to 0.555. GPT-OSS-120B and Qwen3-80B-Thinking perform similarly at 56.2\% and 55.8\% respectively, both achieving the highest FKT accuracy among evaluated models. The reasoning chains appear to provide modest benefit but do not substantially outperform the baseline.

\subsubsection{Cognitive Modeling: What Will Students Write?}

Predicting exact student responses is substantially harder than binary correctness prediction, as expected. Cognitive modeling accuracy, where the model must predict the specific answer a student will produce, ranges from 35.5\% to 44.3\% across models. Llama-3.3-70B achieves the highest cognitive accuracy, though this remains far below what practical student modeling would require.

Table~\ref{tab:kt-by-problem-type} breaks down cognitive accuracy by problem type.

\begin{table}[h]
\centering
\caption{Cognitive modeling accuracy by problem type.}
\label{tab:kt-by-problem-type}
\begin{tabular}{lccc}
\toprule
\textbf{Model} & \textbf{MC (select 1)} & \textbf{MC (select all)} & \textbf{Fill-in-blank} \\
\midrule
GPT-OSS-120B & 44.2\% & \textbf{14.7\%} & 45.6\% \\
Llama-3.3-70B-Instruct & \textbf{47.0\%} & 12.6\% & \textbf{49.9\%} \\
Qwen3-Next-80B-Instruct & 37.6\% & 12.6\% & 39.4\% \\
Qwen3-Next-80B-Thinking & 43.2\% & 13.1\% & 42.2\% \\
\midrule
\textit{Random Baseline} & $\sim$41.3\% & $\sim$2.9\% & $\sim$0\% \\
\bottomrule
\end{tabular}
\end{table}

Multi-select problems prove extremely difficult. All models achieve only 12-15\% accuracy on ``select all'' problems, exceeding the random baseline of approximately 2.9\% but not performing reliably well. Predicting the exact subset of options a student will choose is combinatorially challenging.

For multiple choice problems, all models besides Qwen-3-Instruct outperformed guessing, but not by much. This suggests LLMs may not be able to predict which distractors are truly modelling student misconceptions. However, a limitation of \datasetnamenospace{} is that while almost all images in problems have alt-text, there are several problems with answer-options that do not include alt-text, limiting the benefits of NLP from LLMs, which means the true baseline is slightly below 41.3\%. 

The fill-in-the-blank performance of 40-50\% is particularly notable: for open-ended numerical responses where random chance is essentially zero, models demonstrate genuine capability to anticipate what students will write. While these responses are typically just a few numbers, fractions, or short algebraic expressions, this suggests some understanding of common student responses, even if overall knowledge tracing performance disappoints. However, looking at the cognitive modelling for correct and incorrect answers tells a more compelling story.

\subsubsection{The Optimism Bias}

Table~\ref{tab:kt-bias} reveals a pattern that has significant implications for tutoring applications.

\begin{table}[h]
\centering
\caption{Knowledge tracing accuracy broken down by correctness.}
\label{tab:kt-bias}
\begin{tabular}{lccccc}
\toprule
& \multicolumn{2}{c}{\textbf{When Student Correct}} & \multicolumn{2}{c}{\textbf{When Student Incorrect}} & \textbf{When Answer Incorrect} \\
\cmidrule(lr){2-3} \cmidrule(lr){4-5} \cmidrule(lr){6-6}
\textbf{Model} & \textbf{FKT Acc.} & \textbf{Cog. Acc.} & \textbf{FKT Acc.} & \textbf{Cog. Acc.} & \textbf{Cog. Acc.} \\
\midrule
GPT-OSS-120B & 68.2\% & 62.9\% & 43.4\% & 18.2\% & 10.1\% \\
Llama-3.3-70B & \textbf{85.4\%} & \textbf{72.6\%} & 12.6\% & 14.5\% & 3.7\% \\
Qwen3-80B-Inst. & 56.5\% & 50.3\% & \textbf{51.0\%} & \textbf{19.8\%} & \textbf{13.5\%} \\
Qwen3-80B-Think. & 63.6\% & 57.6\% & 44.6\% & 18.3\% & 12.9\% \\
\bottomrule
\end{tabular}
\end{table}

All models perform far better when students actually answer correctly than when they answer incorrectly. The most extreme case is Llama-3.3-70B: it achieves 85.4\% accuracy when the ground truth is correct but only 12.6\% when incorrect. The model is heavily biased toward predicting that students would answer correctly.

This discrepancy is problematic for educational applications. An effective tutor must recognize when students are struggling, intervening before misconceptions become ingrained and students become frustrated. A model that defaults to optimism, that consistently predicts students are performing well, would systematically fail to provide help when students need it most. The bias may stem from LLMs' pretraining on correct reasoning patterns; models have learned to produce right answers, not to anticipate wrong ones.

Qwen3-Next-80B-Instruct shows the smallest gap between correct-case and incorrect-case performance (56.5\% vs. 51.0\%), though at the cost of lower overall accuracy. This more calibrated behavior, while imperfect, may be preferable for applications that require reliable identification of struggling students.

Students can input the correct answer on their first try, and still be incorrect, if they requested a hint or requested to see the answer. As can be seen in the rightmost column of table \ref{tab:kt-bias}, cognitive accuracy drops to 13.5\% for Qwen3 Instruct, and just 3.7\% when the student's answer does not match the correct answer. Compared to table \ref{tab:kt-by-problem-type}, this demonstrates that predicting the exact answer a student will give remains an open problem, particularly for incorrect answers.

\subsubsection{More Context Does Not Help}

\begin{figure}[t]
    \centering
    \includegraphics[width=\textwidth]{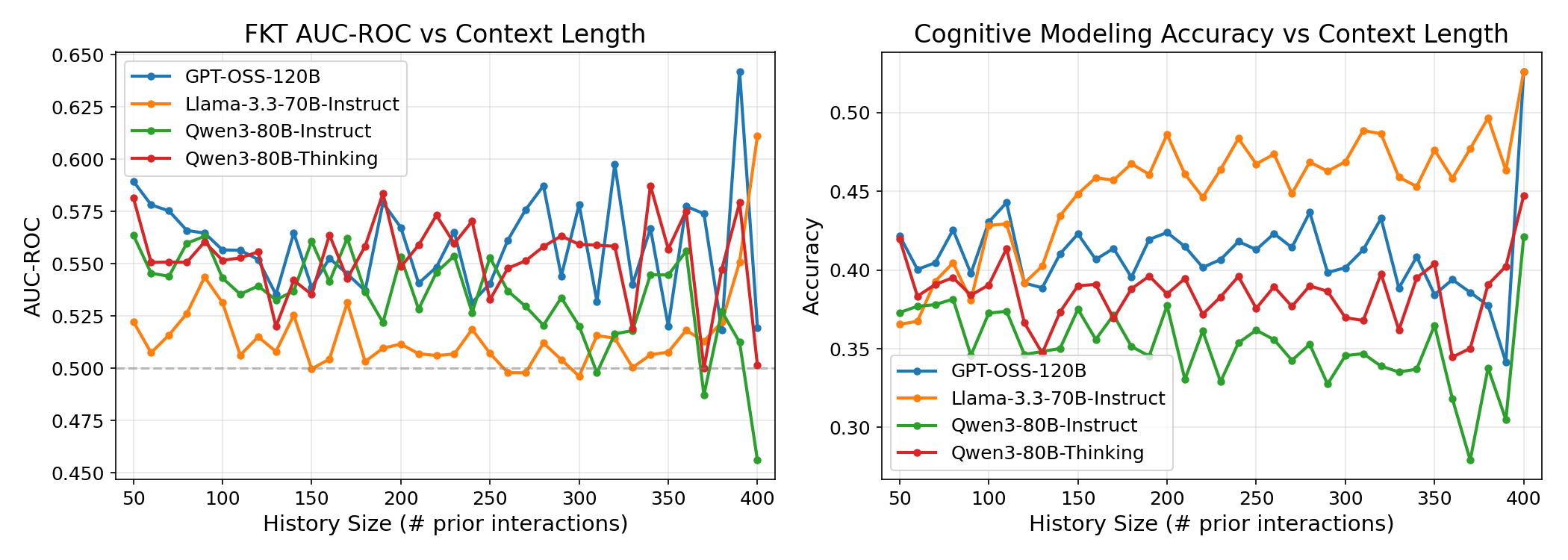}
    \caption{KT performance vs context length (50--400 prior interactions). Left: FKT AUC-ROC remains near random baseline (0.5) regardless of history size. Right: Cognitive modeling accuracy shows no consistent improvement with longer context.}
    \label{fig:kt-context-scaling}
\end{figure}

A natural hypothesis is that models might perform better with longer student histories, allowing them to use more context to identify learning patterns. Figure~\ref{fig:kt-context-scaling} tests this hypothesis by evaluating performance across history sizes from 50 to 400 prior interactions.

The results show context length has essentially no effect on FKT performance. All models hover around AUC-ROC of 0.5--0.6 regardless of whether they observe 50 or 400 prior student interactions. This suggests that LLMs are not effectively leveraging the additional context to build better student models. Either the relevant signal for predicting future performance is already present in shorter histories, or the models cannot extract meaningful patterns from student interaction sequences at longer lengths.

Cognitive modeling shows similarly flat trends, with the notable exception of Llama-3.3-70B, which improves from 37\% to 48\% accuracy as context grows. However, as Llama typically outputs the correct answer, it is unlikely it is learning something anything useful about students' cognitive models.

\subsubsection{Summary}

Knowledge tracing remains a significant challenge for current LLMs. All models barely achieve a trivial prior baseline, suggesting that despite substantial context and reasoning capability, these models cannot effectively leverage student histories for prediction. Reasoning models appear to perform slightly better, albeit marginally. Cognitive modeling shows some capability to anticipate student responses, but this is much more prevalent for correct answers. The systematic optimistic bias is perhaps most concerning: models that cannot recognize when students will fail are poorly suited for the adaptive intervention that personalized education requires.

\subsection{Pedagogical Grounding Results}
\label{sec:pg-results}

\subsubsection{Establishing Ground Truth}

Before presenting results, we describe how ground truth was computed for each pedagogical grounding task.

\begin{figure}[t]
    \centering
    \includegraphics[width=\textwidth]{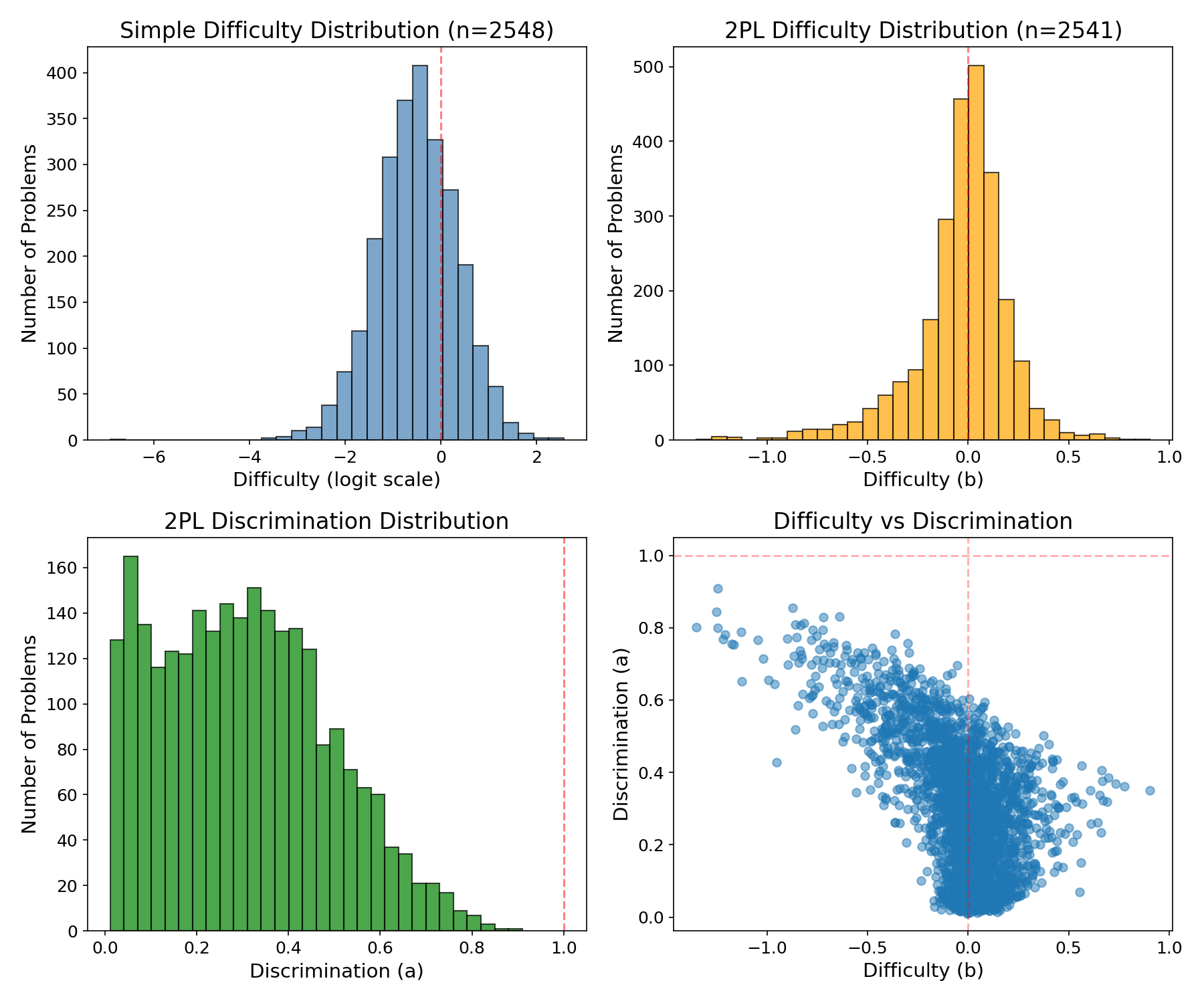}
    \caption{Distribution of IRT parameters across 2,548 problems. Difficulty ranges from -1.35 (easy) to 0.91 (hard); discrimination ranges from 0.01 (uninformative) to 0.91 (highly diagnostic).}
    \label{fig:irt-params}
\end{figure}

For IRT parameters, we computed difficulty and discrimination using 2-parameter Bayesian IRT (py-irt) on the full dataset of 1.7 million student responses. After filtering for problems with at least 50 responses, we obtained parameters for 2,548 problems. Figure~\ref{fig:irt-params} shows the resulting distributions: difficulty ranges from -1.35 (easy) to 0.91 (hard), while discrimination ranges from 0.01 (uninformative) to 0.91 (highly diagnostic).

\begin{figure}[t]
    \centering
    \includegraphics[width=\textwidth]{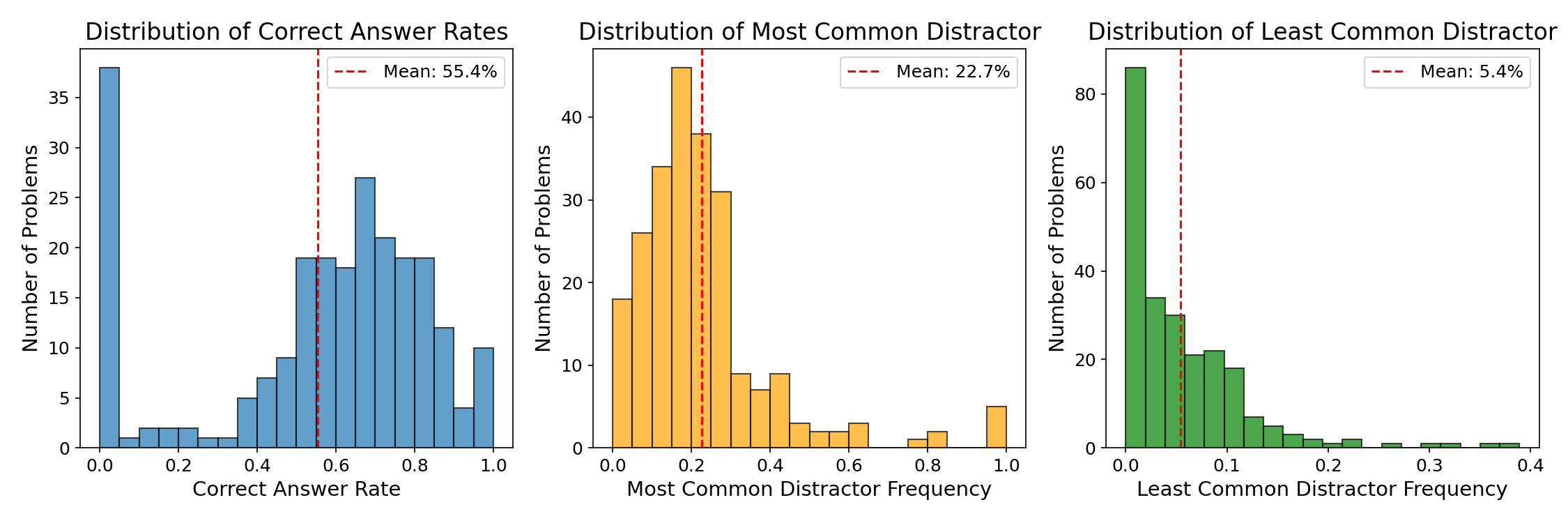}
    \caption{Distractor selection frequencies across 236 multiple-choice problems. Some distractors capture common misconceptions (high selection), while others are rarely chosen (low effectiveness).}
    \label{fig:distractor-stats}
\end{figure}

For distractor statistics, we analyzed 236 multiple-choice problems, computing the empirical selection frequency of each distractor from the student responses. Figure~\ref{fig:distractor-stats} shows the distribution: some distractors capture common misconceptions and are frequently selected, while others are rarely chosen. The most and least commonly selected distractors serve as ground truth for the distractor prediction tasks.

\subsubsection{Difficulty: Moderate Understanding}

Table~\ref{tab:pg-main-results} presents the main pedagogical grounding results.

\begin{table}[h]
\centering
\caption{Pedagogical Grounding results. All tasks are zero-shot.}
\label{tab:pg-main-results}
\begin{tabular}{lcccc}
\toprule
\textbf{Model} & \textbf{Difficulty} & \textbf{Discrimination} & \textbf{Distractor Most} & \textbf{Distractor Least} \\
\midrule
GPT-OSS-120B & \textbf{68.6\%} & 31.1\% & 35.5\% & 31.2\% \\
Llama-3.3-70B & 65.7\% & 40.3\% & 43.6\% & 28.2\% \\
Qwen3-80B-Instruct & 63.9\% & 43.5\% & \textbf{47.9\%} & \textbf{39.7\%} \\
Qwen3-80B-Thinking & 63.7\% & \textbf{46.9\%} & 40.2\% & 20.5\% \\
\midrule
\textit{Random Baseline} & 50.0\% & 50.0\% & 35.8\% & 35.8\% \\
\bottomrule
\end{tabular}
\end{table}

On difficulty comparison, all models substantially outperform random chance. GPT-OSS-120B achieves the best accuracy at 68.6\%, exceeding the 50\% baseline by 18.6 percentage points. This indicates that LLMs have an internalized understanding of what makes mathematical problems hard, potentially from exposure to difficulty-tagged educational content during pretraining.

Table~\ref{tab:pg-difficulty-stratum} examines how performance varies with the magnitude of difficulty differences.

\begin{table}[h]
\centering
\caption{Difficulty comparison accuracy by stratum. Larger differences are easier to detect.}
\label{tab:pg-difficulty-stratum}
\begin{tabular}{lccc}
\toprule
\textbf{Model} & \textbf{Small (0.1--0.5)} & \textbf{Medium (0.5--1.0)} & \textbf{Large ($>$1.0)} \\
\midrule
GPT-OSS-120B & \textbf{57.2\%} & \textbf{68.5\%} & \textbf{80.0\%} \\
Llama-3.3-70B & 57.0\% & 66.2\% & 74.0\% \\
Qwen3-80B-Instruct & 54.9\% & 62.9\% & 74.0\% \\
Qwen3-80B-Thinking & 54.6\% & 62.7\% & 73.8\% \\
\bottomrule
\end{tabular}
\end{table}

The pattern is intuitive: models perform better when difficulty differences are larger. GPT-OSS-120B achieves 80\% accuracy when IRT difficulty gaps exceed 1.0, but only 57.2\% for gaps between 0.1 and 0.5. LLMs can recognize obvious difficulty disparities but struggle with nuanced comparisons.

\subsubsection{Discrimination: A Gap in Knowledge for LLMs}

A concerning finding in our evaluation is that every model performs \textit{below random chance} on discrimination comparison. Even Qwen3-80B-Thinking, the best performer, achieves only 46.9\%, worse than coin flipping.

Table~\ref{tab:pg-lift} quantifies each model's improvement or decline relative to random guessing.

\begin{table}[h]
\centering
\caption{Lift over random baseline (positive = better than random, negative = worse than random).}
\label{tab:pg-lift}
\begin{tabular}{lcccc}
\toprule
\textbf{Model} & \textbf{Difficulty} & \textbf{Discrimination} & \textbf{Distr. Most} & \textbf{Distr. Least} \\
\midrule
GPT-OSS-120B & \textbf{+18.6\%} & -18.9\% & -0.3\% & -4.6\% \\
Llama-3.3-70B & +15.7\% & -9.7\% & +7.8\% & -7.6\% \\
Qwen3-80B-Instruct & +13.9\% & -6.5\% & \textbf{+12.1\%} & \textbf{+4.0\%} \\
Qwen3-80B-Thinking & +13.7\% & \textbf{-3.1\%} & +4.4\% & -15.2\% \\
\bottomrule
\end{tabular}
\end{table}

This is not a marginal failure. Discrimination measures how effectively an assessment item separates high-ability from low-ability students. A high-discrimination item produces clear separation: capable students succeed, struggling students fail. A low-discrimination item shows similar performance across ability levels, often because guessing, unclear wording, or irrelevant knowledge requirements confound the measurement. Understanding discrimination is central to educational assessment, and the fact that LLMs perform below chance suggests they do not grasp this concept and somehow have learned the opposite of what discrimination really is.

Table~\ref{tab:pg-discrimination-stratum} examines discrimination accuracy by stratum.

\begin{table}[h]
\centering
\caption{Discrimination comparison accuracy by stratum.}
\label{tab:pg-discrimination-stratum}
\begin{tabular}{lcc}
\toprule
\textbf{Model} & \textbf{Small (0.1--0.5)} & \textbf{Medium (0.5--1.0)} \\
\midrule
GPT-OSS-120B & 38.3\% & 23.9\% \\
Llama-3.3-70B & 44.6\% & 35.9\% \\
Qwen3-80B-Instruct & 46.4\% & 40.7\% \\
Qwen3-80B-Thinking & 48.8\% & 44.9\% \\
\bottomrule
\end{tabular}
\end{table}

The pattern observed here is paradoxical. Unlike difficulty, where larger differences are easier to detect, discrimination accuracy is actually \textit{higher} for smaller differences. This counterintuitive finding points to systematic errors on medium-difference pairs. We hypothesize that models conflate difficulty with discrimination, mistakenly assuming that harder problems are inherently more discriminating. Since difficulty and discrimination are often inversely related, a very hard problem that nearly all students get wrong exhibits low discrimination because it fails to meaningfully separate student performance.

\subsubsection{Why Is Discrimination So Hard?}

Understanding discrimination requires reasoning about how different students would approach a problem. A high-discrimination item has wrong answers that specifically trap less capable students, while capable students avoid them. Recognizing this requires understanding the wide range of possible student errors and how those errors correlate with ability.

This is essentially metacognitive reasoning about learning. It asks not ``what is the right answer?'' but ``on which problems do weak students make mistakes that strong students avoid?'' Such reasoning may be absent from typical LLM pretraining, which focuses on producing correct outputs rather than modeling the distribution of human errors.

\subsubsection{Distractor Prediction: Testing Knowledge of Student Misconceptions}

These tasks test a capability that research has shown matters for student learning. Chen et al.\ found that teachers who can predict which wrong answer students will most commonly choose produce greater learning gains in their students \cite{chen2020impact}. This \textit{Knowledge of Student Misconceptions} (KOSM) is distinct from subject matter expertise: knowing the right answer does not imply understanding which errors students find tempting. If LLMs are to serve as effective educational tools for tasks such as problem and distractor generation, they should demonstrate similar understanding.

\paragraph{Most Common Distractor.}
For predicting which wrong answer students choose most often, results show moderate success. Qwen3-80B-Instruct achieves 47.9\%, exceeding the 35.8\% random baseline by 12.1 percentage points. Llama-3.3-70B reaches 43.6\%, also substantially above chance. These results suggest LLMs have internalized some understanding of common student errors, possibly from exposure to educational content that discusses misconceptions.

Performance varies with the number of distractors (see Appendix Table~\ref{tab:app-distractor-options}). With only 2 distractors, Llama-3.3-70B achieves 65.2\%, far exceeding the 50\% baseline. With 3 distractors, Qwen3-80B-Instruct leads at 50.3\% versus 33.3\% baseline. With 4 distractors, only Qwen3-80B-Thinking achieves the baseline for predicting the most common distractor. As the number of distractors increases, the task becomes harder and performance approaches baseline.

\paragraph{Least Common Distractor.}
Predicting which distractor students rarely choose proves a substantially harder task for our LLMs. Most models perform at or below random chance. Qwen3-80B-Thinking achieves an overall accuracy of only 20.5\%, falling 15.2 percentage points below the 35.8\% baseline. This suggests LLMs can identify which errors are common but struggle to recognize which are implausible.

One interpretation is that LLMs have learned to recognize prevalent misconceptions through pattern matching on educational content, but lack the deeper understanding needed to identify why certain errors are rare. A distractor may be rarely chosen because it is mathematically implausible, because it does not correspond to any coherent reasoning path, or because its surface features signal ``obviously wrong.'' Reasoning about these distinctions requires understanding student cognition in ways that may exceed current LLM capabilities.

\paragraph{Implications for Educational AI.}
The partial success on most-common distractor prediction is encouraging for applications like automated distractor generation and misconception detection. However, the failure on least-common prediction suggests that LLMs cannot yet reliably distinguish mildly effective from ineffective distractors. An LLM-generated distractor might target a common error or might be implausible to students; without KOSM, the system cannot tell the difference.

\subsubsection{Extended Reasoning: A Double-Edged Sword}

The Thinking model's performance reveals an interesting pattern. Reasoning helps with more abstract comparisons: discrimination accuracy improves from 43.5\% (Instruct) to 46.9\% (Thinking), a gain of 3.4 percentage points. But it hurts concrete predictions: most-common distractor accuracy drops from 47.9\% to 40.2\%, a loss of 7.7 percentage points.

Reasoning chains may enable more careful analysis of abstract pedagogical properties while introducing overthinking on tasks that benefit from pattern matching. When predicting which distractor students commonly choose, intuitive responses based on surface-level similarity to common errors may outperform deliberate reasoning about student cognition.

\subsubsection{No Single Best Model}

Table~\ref{tab:pg-lift} reveals that no single model dominates across all tasks. GPT-OSS-120B excels at difficulty comparison but performs worst on discrimination. Qwen3-80B-Instruct is the most balanced, outperforming the random-chance baseline on 3 of 4 tasks. The Thinking variant helps discrimination but hurts distractor prediction. These trade-offs suggest that different pedagogical reasoning tasks may require different capabilities. Success at one does not predict success at others.

\subsubsection{Summary}

Pedagogical grounding results reveal a nuanced picture of LLM capabilities. Models demonstrate genuine understanding of problem difficulty, beating random chance by 14-19 percentage points on difficulty comparison. But they completely fail on discrimination, performing below random and revealing a fundamental gap in understanding how assessment items interact with student ability distributions. Distractor prediction is mixed, with some success at identifying the most frequently chosen distractors, but poor performance on identifying rarer ones. Extended reasoning helps some tasks and hurts others. The discrimination failure is particularly significant. Without understanding what makes one problem more diagnostic than another, LLMs cannot support principled assessment design, which requires identifying which problems provide the most useful information about student knowledge.

\section{Discussion}
\label{sec:discussion}

\subsection{Why Do LLMs Struggle with Educational Tasks?}

Our results reveal systematic limitations in current LLMs for educational applications. Understanding why these limitations exist may guide future research.

\paragraph{Training Data Mismatch.}
LLMs are predominantly trained on text that represents correct reasoning: textbooks present solutions, explanations walk through right answers, and educational content generally models successful problem-solving. Student errors, misconceptions, and learning trajectories are likely underrepresented. This asymmetry may explain the optimistic bias we observe in knowledge tracing: models default to predicting correct answers because that is overwhelmingly what they have learned to produce. Modeling student failures requires exposure to the patterns of student failures, and such data is rare in typical large-scale pre-training corpora.

\paragraph{Missing Student Representations.}
Effective knowledge tracing requires maintaining and updating a latent model of the student that evolves with each interaction, capturing both mastery and misconceptions over time. While autoregressive LLMs can understand, in a natural language sense, long interaction histories, they lack explicit mechanisms for representing or updating student-specific state. Their generative architecture is optimized for modeling surface-level linguistic patterns rather than inferring latent cognitive variables from sequences of student responses. Moreover, pre-training data and objectives do not expose LLMs to the kind of student-modeling supervision required for robust knowledge tracing. As a result, simply increasing context length is insufficient, and effective knowledge tracing may require fine-tuning with educational interaction data or architectural extensions that explicitly represent and update student state.

\paragraph{Lacking Longitudinal Capabilities.}
Compared to models such as Bayesian Knowledge Tracing (BKT), Logistic Knowledge Tracing (LKT), and Deep Knowledge Tracing (DKT), LLMs face additional limitations in leveraging information across multiple students. Traditional KT models are trained on datasets containing responses from tens, hundreds or thousands of students completing the same problem, allowing them to learn statistical patterns of mastery and error across a population. In contrast, when applied to a student-specific setting as in this work, LLMs observe only a single student’s interaction history, with no mechanism for aggregating experience from other students. While it is theoretically possible to provide multiple “shots” of student interactions to LLMs to simulate exposure to a broader dataset, this approach both substantially increases inference time and rapidly consumes the fixed context window of commercially available models. As a result, LLMs cannot naturally benefit from population-level patterns in the same way that BKT, LKT, and DKT do, limiting their effectiveness for longitudinal knowledge tracing. It remains an open question whether the boon of receiving and understanding the textual content of problems and student answers can mitigate this limitation.

\paragraph{Discrimination Requires Metacognition.}
Understanding item discrimination requires reasoning about how \textit{different students} approach a problem. A high-discrimination item elicits different responses from high-ability and low-ability students; recognizing this requires understanding not just what the right answer is, but what errors different types of students would make and why. This is essentially metacognitive reasoning about learning. LLMs that have not been explicitly trained on educational measurement concepts may lack the representational intuition for such reasoning.

\subsection{Implications for Educational AI}

\paragraph{Personalized Tutoring.}
Our findings suggest caution is needed in deploying LLMs for personalized tutoring. A tutor must recognize when students are struggling to provide timely interventions. However, our results show models that cannot accurately predict when students will struggle, as is exhibited by the FKT accuracy discrepancy between correct and incorrect answers. Each exhibits strong optimistic bias, predicting success when students will fail. An LLM tutor with these characteristics would systematically miss opportunities to help students when they need it most.

\paragraph{Assessment Design.}
The difficulty comparison results (up to 68.6\% accuracy) suggest LLMs could assist in sorting problems by difficulty. This capability might help content developers organize practice materials or estimate item characteristics before running costly experiments to verify with empirical results. However, the complete failure of discrimination indicates that LLMs should not be relied upon for designing high-stakes assessments without human oversight. Selecting items that effectively measure student ability requires understanding discrimination, and current LLMs demonstrably lack this understanding. Estimating difficulty, and to a lesser extent discrimination, may be more feasible for LLMs than full knowledge tracing, as these tasks align better with the natural language understanding that LLMs are typically trained and calibrated for.

\paragraph{Training Data Needs.}
Our results highlight the need for LLM training data that includes student learning processes, not just correct solutions. Models exposed only to successful reasoning cannot be expected to anticipate unsuccessful reasoning. \datasetnamenospace{} itself could serve as a resource for developing such training: it provides extensive examples of student errors, misconceptions, and learning trajectories that are typically absent from pretraining data.

\subsection{Limitations}

\paragraph{Domain Scope.}
\datasetnamenospace{} focuses exclusively on K-12 mathematics. Results may not generalize to other subjects such as language arts or science, or to other educational levels such as higher education. Mathematics has particular characteristics, including clear right/wrong answers and hierarchical skill dependencies, that may make it atypical of educational domains generally.

\paragraph{Model Selection.}
We evaluated four models from three organizations, sampling the frontier of current LLM capabilities. However, the LLM landscape is evolving rapidly. Future models may address some of the limitations we have identified, particularly if trained with educational objectives in mind. Our results establish baselines against which progress can be measured.

\paragraph{Zero-Shot Evaluation.}
All evaluations used zero-shot prompting. Fine-tuning on educational data, or few-shot prompting with carefully chosen examples, might substantially improve performance. However, such evaluations would test different capabilities: not what LLMs understand from pretraining, but what they can learn from targeted supervision. Our results characterize the educational understanding that emerges from general-purpose pretraining.

\paragraph{IRT Ground Truth.}
IRT parameters are estimated with associated uncertainty. Discrimination parameters in particular can be noisy for items with limited response data. While we filtered for items with at least 50 responses, ground truth labels inevitably contain measurement error that propagates into evaluation.

\paragraph{Distractor Sample Size.}
The distractor prediction tasks use only 236 multiple-choice problems, limiting statistical power for fine-grained analysis. Results on these tasks should be interpreted with appropriate caution.

\paragraph{Changes to ASSISTments.} Problems in ASSISTments went through changes during the time the data in \datasetnamenospace{} was collected. A small number of problems changed problem types, which was not recorded. Further, there are a limited number of problems on which the correct answer was mistyped when first added to the platform. Accordingly, there are a limited number of interactions where students' responses or correctness may not be aligned with what they would currently have received.

\paragraph{Environmental Concerns.}
Large language models (LLMs) require substantial energy for pretraining, fine-tuning, and inference. Compared to traditional knowledge tracing methods such as BKT and DKT, the environmental footprint of LLMs is considerably higher. While we believe that advancing educational applications can justify the use of LLMs, their energy consumption is non-negligible and should be weighed carefully, especially in cases where they offer only marginal gains over more efficient algorithms.

\subsection{Broader Impact}

\paragraph{Positive Impacts.}
By establishing clear benchmarks and identifying specific failure modes, we aim to guide the development of more capable educational AI. Researchers now know that with LLMs discrimination understanding is a gap requiring attention, that optimistic bias is a practical concern, and that reliable knowledge tracing remains unsolved. The dataset release enables reproducible research and fair comparison of future methods against current baselines.

\paragraph{Risks.}
Overconfidence in LLM educational capabilities could lead to premature deployment, potentially harming students. Our results show substantial room for improvement before LLMs should be trusted for high-stakes educational decisions. We emphasize that current models are not suitable for applications requiring reliable student modeling or assessment quality judgments.

\paragraph{Equity Considerations.}
Educational AI systems must work equitably across student populations. \textbf{\datasetnamenospace{}} does not include results by student demographics or prior achievement levels. Future work should examine whether model performance varies systematically across student subgroups, identifying potential disparities before deployment.

\section{Conclusion}
\label{sec:conclusion}

We opened with a question: can Large Language Models understand how students learn? After comprehensive evaluation on \textbf{\datasetnamenospace{}}, the first English educational dataset that provides the textual richness LLMs require, our findings are clear: significant gaps remain. Current frontier models possess fragments of educational understanding, but they lack the integrated capabilities needed for reliable personalized instruction.

\datasetnamenospace{} itself represents a significant contribution to educational AI research. Prior knowledge tracing benchmarks offered only question identifiers and binary correctness labels, which tell an LLM nothing about what makes problems hard or which errors reveal misconceptions. Our dataset provides what has been missing: complete question text for 3,400 mathematics problems, actual student responses for 1.7 million interactions, records of which distractors students selected, and alignment to Common Core K-12 standards. These 5,000 students with their full learning trajectories enable research questions that were previously impossible to investigate. We will release the dataset publicly with the support of ASSISTments.

Our six benchmark tasks evaluate complementary facets of educational understanding. The knowledge tracing track revealed that all evaluated models barely achieve a trivial baseline, achieving at best 56.2\% accuracy against the 51.3\% baseline probability. The pedagogical grounding track uncovered a more fundamental limitation: every model performs below random chance on discrimination comparison, suggesting that LLMs do not understand what makes one assessment item more diagnostic than another. Models do show competence at judging relative difficulty, with accuracy reaching 80.0\% for substantial difficulty gaps, and some success at predicting common distractors. These partial successes make the failures more intriguing--difficulty is learnable from pretraining data, but what is needed for a more thorough understanding of discrimination remains unclear.

Several research directions emerge from these findings. Fine-tuning on educational data may help models acquire the student modeling capabilities absent from general pretraining. The challenge of long student histories, which strain context limits, invites work on selective attention to pedagogically relevant interactions. And the discrimination failure in particular may require a new approach to understanding the relationship between the natural language of questions and the pitfalls encountered by low- and high-achieving students.

We hope \datasetnamenospace{} catalyzes this research, just as the original ASSISTments datasets have served as a benchmark in over a decade's worth of research in knowledge tracing. By providing a large, natural language-oriented benchmark with a variety of potential educational tasks on which this dataset could be used, researchers will be able to determine where and how LLMs can be effectively implemented into educational settings.

\section*{Acknowledgments}
We thank the ASSISTments team for providing the foundational data that made this research possible.
We thank our many past and current funders, including:
NSF (2118725, 2118904, 1950683, 1917808, 1931523, 1940236, 1917713, 1903304, 1822830, 1759229, 1724889, 1636782, 1535428, 2215842, 2225091, 2341948, \& 2153481); 
IES (R305N210049, R305D210031, R305A170137, R305A170243, R305A180401, R305A120125, R305R220012, \& R305T240029); 
GAANN (P200A120238, P200A180088, \& P200A150306); DOE (U411B190024 S411B210024, S411B220024, \& S411A240012 ); ONR (N00014-18-1-2768); NIH (via a SBIR R44GM146483), Schmidt Futures, BMGF, CZI, Arnold, Hewlett, the Jaffe Foundation, and anonymous donors. None of the opinions expressed here are those of the funders.

\bibliographystyle{unsrt}
\bibliography{refs}

\clearpage
\appendix

\section{Prompt Templates}
\label{app:prompts}

This appendix provides the complete prompt templates used for all benchmark tasks.

\subsection{Knowledge Tracing Prompts}

\subsubsection{System Prompt}

For knowledge tracing, we predicted skill-level mastery, question-level correctness, and exact answer. Due to the ability of LLMs to understand surface-level constructs, such as problem correctness, but not latent-level constructs, such as skill-mastery, we only included question-level correctness and exact answer predictions in our task evaluations. The following system prompt is used for all Knowledge Tracing tasks:

\begin{quote}
\small
\ttfamily
You are a reasoning model trained to simulate a student's evolving knowledge and response behavior in mathematics.

Your goal is to infer, from past problem--answer pairs, how this same student is likely to perform on a new problem --- at multiple levels of granularity.

You must reason about the student's learning progression, skill mastery, and recurring misconceptions, then produce structured predictions for the new item.

---

Your Task:

Generate three coordinated predictions for this student:

1) **Skill-level knowledge tracing (0 or 1):** Whether the student has mastered the underlying skill involved in the new problem.

2) **Question-level knowledge tracing (0 or 1):** Whether the student will answer this specific problem correctly.

3) **Cognitive-level prediction (string):** The exact answer text or option the student would most likely produce, written in their own response style.

---

Reasoning Guidelines:

- Use the student's historical data (problems, answers, hints, timestamps) to infer learning and forgetting patterns.

- Consider recency and exposure: later timestamps often indicate updated knowledge.

- Treat UsedHint=True or SawAnswer=True as evidence that the student's recorded answer may not reflect true mastery.

- Attend to how the student's accuracy, style, and misconceptions evolve over time.

---

Output Format:

Finish your response with a JSON object:

For Multiple Choice (select 1) problems:\\
\{"skill\_level": 0 or 1, "question\_level": 0 or 1, "student\_answer": "A"\}

For Multiple Choice (select all) problems:\\
\{"skill\_level": 0 or 1, "question\_level": 0 or 1, "student\_answer": "A, C"\}

For Fill-in problems:\\
\{"skill\_level": 0 or 1, "question\_level": 0 or 1, "student\_answer": "<value>"\}
\end{quote}

\subsubsection{User Prompt Template}

The user prompt presents the student's history and the new problem:

\begin{quote}
\small
\ttfamily
You will receive a series of the student's prior problems and responses.\\

For each history item:\\
- ProblemID: <id>\\
- Timestamp: <timestamp>\\
- Problem: <problem text>\\
- Problem Type: Multiple Choice / Fill-in\\
- Options: A) ... B) ... C) ...\\
- Correct Answer: <letter or value>\\
- Student's First Answer: <letter or value>\\
- UsedHint: <True/False>\\
- SawAnswer: <True/False>\\
- Skill: <skill\_name>\\
\\
**Student's Previous Problems:**\\
\indent \textit{[History items formatted as above]}\\
\\
**New Problem to Predict:**\\
Timestamp: <timestamp>\\
Problem: <problem text>\\
Problem Type: <type>\\
Answer Options: A) ... B) ... C) ...\\
Skill: <skill\_name>\\
\\
Make your predictions for this student.
\end{quote}

\subsection{Pedagogical Grounding Prompts}

\subsubsection{Difficulty Comparison}

\textbf{System Prompt:}
\begin{quote}
\small
\ttfamily
You are an expert educator evaluating mathematics problems for instructional design.

Your task is to compare two problems and determine which one is MORE DIFFICULT.

A problem is MORE DIFFICULT if:
- It requires more prerequisite knowledge or skills
- It involves more complex reasoning steps
- It has higher cognitive load
- Students are more likely to make errors
- It requires synthesizing multiple concepts

Instructions:
1. Carefully read both Problem A and Problem B
2. Consider factors like conceptual complexity, prerequisite knowledge, cognitive load, and potential for errors
3. Make your judgment based solely on the problem content
4. Respond with exactly "A" or "B"

Output Format:\\
Respond with a single JSON object:\\
\{"answer": "A"\} or \{"answer": "B"\}
\end{quote}

\textbf{User Prompt Template:}
\begin{quote}
\small
\ttfamily
Compare the following two problems and determine which is more difficult.

**Problem A:**\\{}
[Problem A text]

**Problem B:**\\{}
[Problem B text]

Which problem is more difficult? Respond with \{"answer": "A"\} or \{"answer": "B"\}.
\end{quote}

\subsubsection{Discrimination Comparison}

\textbf{System Prompt:}
\begin{quote}
\small
\ttfamily
You are an expert educator evaluating mathematics problems for instructional design.

Your task is to compare two problems and determine which one has HIGHER DISCRIMINATION.

A problem has HIGHER DISCRIMINATION if:
- It better distinguishes between students who understand vs. those who don't
- Correct answers strongly indicate mastery
- Incorrect answers are unlikely for knowledgeable students
- The problem tests specific, well-defined skills
- It avoids ambiguity that could confuse strong students

Instructions:
1. Carefully read both Problem A and Problem B
2. Consider how well each problem separates students by ability level
3. Make your judgment based solely on the problem content
4. Respond with exactly "A" or "B"

Output Format:\\
Respond with a single JSON object:\\
\{"answer": "A"\} or \{"answer": "B"\}
\end{quote}

\subsubsection{Most Common Distractor}

\textbf{System Prompt:}
\begin{quote}
\small
\ttfamily
You are an expert educator analyzing student misconceptions in mathematics.

Your task is to predict which incorrect answer option (distractor) is MOST COMMONLY CHOSEN by students.

The MOST COMMON distractor is the wrong answer that students choose most frequently.
Common reasons include:
- It represents a predictable computational error
- It matches a common misconception
- It seems plausible but contains a subtle flaw
- It results from a partial understanding of the concept
- It's what you get if you make a typical arithmetic mistake

Instructions:
1. Read the problem carefully
2. Identify the correct answer
3. Consider common student misconceptions and errors
4. Analyze each incorrect answer option
5. Predict which distractor students would most often choose
6. Respond with the letter of that distractor

Output Format:\\
Respond with a single JSON object:\\
\{"answer": "X"\} where X is the letter of the most common distractor
\end{quote}

\textbf{User Prompt Template:}
\begin{quote}
\small
\ttfamily
Analyze this multiple-choice problem and predict which distractor is MOST COMMONLY chosen by students.

**Problem:**\\{}
[Problem text]

**Answer Options:**\\
A) [Option A]\\
B) [Option B]\\
C) [Option C]\\
D) [Option D]

**Correct Answer:** [Letter]

Which incorrect answer option do students choose most often? Respond with \{"answer": "X"\}.
\end{quote}

\subsubsection{Least Common Distractor}

\textbf{System Prompt:}
\begin{quote}
\small
\ttfamily
You are an expert educator analyzing student misconceptions in mathematics.

Your task is to predict which incorrect answer option (distractor) is LEAST COMMONLY CHOSEN by students.

The LEAST COMMON distractor is the wrong answer that students rarely choose.
Common reasons include:
- It is obviously incorrect to most students
- It doesn't correspond to any common error pattern
- It requires a very unusual misconception
- It is far from the correct answer conceptually
- It looks implausible compared to other options

Instructions:
1. Read the problem carefully
2. Identify the correct answer
3. Consider which wrong answer would seem most obviously incorrect
4. Analyze each incorrect answer option
5. Predict which distractor students would rarely choose
6. Respond with the letter of that distractor

Output Format:\\
Respond with a single JSON object:\\
\{"answer": "X"\} where X is the letter of the least common distractor
\end{quote}

\section{Additional Results}
\label{app:results}

\subsection{Extended Knowledge Tracing Results}

Table~\ref{tab:app-kt-extended} provides additional breakdowns of knowledge tracing performance.

\begin{table}[h]
\centering
\caption{Extended Knowledge Tracing results with prediction counts and standard errors.}
\label{tab:app-kt-extended}
\begin{tabular}{lcccccc}
\toprule
\textbf{Model} & \textbf{Total N} & \textbf{Valid FKT} & \textbf{FKT Acc} & \textbf{SE} & \textbf{Cog Acc} & \textbf{SE} \\
\midrule
GPT-OSS-120B & 27,715 & 27,695 & 56.2\% & 0.30\% & 41.2\% & 0.30\% \\
Llama-3.3-70B & 27,715 & 26,457 & 52.3\% & 0.31\% & 44.3\% & 0.30\% \\
Qwen3-80B-Inst. & 27,618 & 27,618 & 54.0\% & 0.30\% & 35.5\% & 0.29\% \\
Qwen3-80B-Think. & 27,715 & 27,014 & 55.8\% & 0.30\% & 38.5\% & 0.30\% \\
\bottomrule
\end{tabular}
\end{table}

\subsection{Cognitive Modeling by Ground Truth and Problem Type}

Table~\ref{tab:app-cognitive-detailed} provides a detailed breakdown of cognitive modeling accuracy across all combinations of ground truth and problem type.

\begin{table}[h]
\centering
\caption{Cognitive modeling accuracy by ground truth (correct/incorrect) and problem type.}
\label{tab:app-cognitive-detailed}
\small
\begin{tabular}{llcccc}
\toprule
\textbf{Ground Truth} & \textbf{Problem Type} & \textbf{GPT-OSS} & \textbf{Llama} & \textbf{Qwen-I} & \textbf{Qwen-T} \\
\midrule
\multirow{3}{*}{Correct} & MC (select 1) & 51.6\% & 65.8\% & 32.3\% & 47.8\% \\
& MC (select all) & 26.0\% & 29.0\% & 25.3\% & 23.7\% \\
& Fill-in-blank & 72.1\% & 81.1\% & 60.3\% & 65.8\% \\
\midrule
\multirow{3}{*}{Incorrect} & MC (select 1) & 30.8\% & 13.3\% & 47.0\% & 35.0\% \\
& MC (select all) & 8.3\% & 3.5\% & 5.4\% & 7.2\% \\
& Fill-in-blank & 18.2\% & 17.6\% & 17.9\% & 17.7\% \\
\bottomrule
\end{tabular}
\end{table}

\subsection{Extended Pedagogical Grounding Results}

Table~\ref{tab:app-pg-extended} provides the full breakdown of pedagogical grounding results.

\begin{table}[h]
\centering
\caption{Extended Pedagogical Grounding results with sample sizes.}
\label{tab:app-pg-extended}
\begin{tabular}{lcccc}
\toprule
\textbf{Task} & \textbf{N} & \textbf{Baseline} & \textbf{Best Model} & \textbf{Best Acc} \\
\midrule
Difficulty & 9,999 & 50.0\% & GPT-OSS-120B & 68.6\% \\
Discrimination & 6,666 & 50.0\% & Qwen3-80B-Think & 46.9\% \\
Distractor Most & 234 & 35.8\% & Qwen3-80B-Inst & 47.9\% \\
Distractor Least & 234 & 35.8\% & Qwen3-80B-Inst & 39.7\% \\
\bottomrule
\end{tabular}
\end{table}

\subsection{Distractor Prediction by Number of Options}

Table~\ref{tab:app-distractor-options} breaks down distractor prediction accuracy by the number of answer options.

\begin{table}[h]
\centering
\caption{Distractor prediction accuracy by number of distractors (N-1 where N = total options).}
\label{tab:app-distractor-options}
\begin{tabular}{lccccc}
\toprule
\textbf{N Distractors} & \textbf{Baseline} & \textbf{GPT-OSS} & \textbf{Llama} & \textbf{Qwen-I} & \textbf{Qwen-T} \\
\midrule
\multicolumn{6}{c}{\textit{Most Common Distractor}} \\
\midrule
2 & 50.0\% & 45.7\% & 65.2\% & 52.2\% & 45.7\% \\
3 & 33.3\% & 34.3\% & 40.2\% & 50.3\% & 40.8\% \\
4 & 25.0\% & 15.4\% & 23.1\% & 15.4\% & 30.8\% \\
\midrule
\multicolumn{6}{c}{\textit{Least Common Distractor}} \\
\midrule
2 & 50.0\% & 50.0\% & 41.3\% & 54.3\% & 39.1\% \\
3 & 33.3\% & 28.4\% & 25.4\% & 36.7\% & 16.6\% \\
4 & 25.0\% & 15.4\% & 30.8\% & 38.5\% & 15.4\% \\
\bottomrule
\end{tabular}
\end{table}

\section{Model Configurations}
\label{app:models}

\subsection{vLLM Inference Settings}

All models were deployed using vLLM with the shared configuration shown in Table~\ref{tab:app-vllm-config}.

\begin{table}[h]
\centering
\caption{Shared vLLM inference configuration.}
\label{tab:app-vllm-config}
\begin{tabular}{ll}
\toprule
\textbf{Parameter} & \textbf{Value} \\
\midrule
Tensor Parallelism & 4 GPUs \\
GPU Memory Utilization & 0.90 \\
Max Model Length & 32,768 tokens \\
Max Num Sequences & 512 \\
Batch Size (KT) & 512 \\
Batch Size (PG) & 500 \\
\bottomrule
\end{tabular}
\end{table}

\subsection{Model-Specific Configurations}

Table~\ref{tab:app-model-configs} lists the generation parameters for each model, following vendor-recommended settings.

\begin{table}[h]
\centering
\caption{Model-specific generation parameters.}
\label{tab:app-model-configs}
\begin{tabular}{lcccc}
\toprule
\textbf{Parameter} & \textbf{GPT-OSS} & \textbf{Llama} & \textbf{Qwen-I} & \textbf{Qwen-T} \\
\midrule
Model ID & gpt-oss-120b & llama-3.3-70b & qwen3-next-80b-i & qwen3-next-80b-t \\
Temperature & 0.7 & 0.7 & 0.7 & 0.6 \\
Top-$p$ & 0.95 & 0.9 & 0.8 & 0.95 \\
Top-$k$ & 20 & --- & 20 & 20 \\
Max Tokens & 32,768 & 32,768 & 32,768 & 32,768 \\
Special Mode & Reasoning prefix & --- & --- & Native thinking \\
\bottomrule
\end{tabular}
\end{table}

\subsection{Hardware and Compute}

Table~\ref{tab:app-compute} summarizes the compute resources used for all experiments.

\begin{table}[h]
\centering
\caption{Compute resources used for experiments.}
\label{tab:app-compute}
\begin{tabular}{ll}
\toprule
\textbf{Resource} & \textbf{Specification} \\
\midrule
GPU Model & NVIDIA A100 80GB \\
GPUs per Experiment & 4--8 \\
CPU & AMD EPYC 7742 (or equivalent) \\
RAM & 512GB \\
Storage & NVMe SSD \\
\midrule
\multicolumn{2}{l}{\textit{Experiment Duration}} \\
KT (per model) & $\sim$48 hours \\
PG (per model) & $\sim$2--4 hours \\
Total GPU-Hours & $\sim$200 \\
\bottomrule
\end{tabular}
\end{table}

\subsection{Reproducibility}

To ensure reproducibility:
\begin{itemize}[leftmargin=*]
    \item All experiments use model-recommended sampling (temperature 0.6--0.7)
    \item Random seeds are fixed (seed=42 for sampling)
    \item All prompts and configurations are provided in this appendix
    \item Evaluation scripts are included in the dataset release
    \item Model checkpoints are publicly available (with access) from respective organizations
\end{itemize}

\subsection{Response Parsing}

Model responses are parsed using the following strategy:
\begin{enumerate}[leftmargin=*]
    \item Search for JSON objects in the response using regex: \texttt{\textbackslash\{[\textasciicircum\{\}]*\textbackslash\}}
    \item Extract the last JSON object found (to handle multi-step reasoning)
    \item Parse JSON and extract relevant fields (\texttt{skill\_level}, \texttt{question\_level}, \texttt{student\_answer} for KT; \texttt{answer} for PG)
    \item If JSON parsing fails, fall back to regex search for standalone letters (A/B/C/D)
    \item If all parsing fails, mark prediction as invalid (counted as incorrect)
\end{enumerate}

Parse failure rates were low ($<$2\%) across all models, with Qwen3-80B-Thinking having slightly higher failure rates due to longer, more complex reasoning chains.

\end{document}